\documentclass[preprint,12pt]{aastex}

\begin{document}

\title{Tracing Sagittarius Structure with SDSS and SEGUE imaging and spectroscopy}

\author{
Brian Yanny\altaffilmark{\ref{FNAL}},
Heidi Jo Newberg\altaffilmark{\ref{RPI}},
Jennifer A. Johnson\altaffilmark{\ref{OSU}},
Young Sun Lee\altaffilmark{\ref{MSU}},
Timothy C. Beers\altaffilmark{\ref{MSU}},
Dmitry Bizyaev\altaffilmark{\ref{APO}},
Howard Brewington\altaffilmark{\ref{APO}},
Paola Re Fiorentin\altaffilmark{\ref{Slovenia},\ref{MPH}},
Paul Harding\altaffilmark{\ref{CASE}},
Elena Malanushenko\altaffilmark{\ref{APO}},
Viktor Malanushenko\altaffilmark{\ref{APO}},
Dan Oravetz\altaffilmark{\ref{APO}},
Kaike Pan\altaffilmark{\ref{APO}},
Audrey Simmons\altaffilmark{\ref{APO}},
Stephanie Snedden\altaffilmark{\ref{APO}}
}

\altaffiltext{1}{Fermi National Accelerator Laboratory, P.O. Box 500, Batavia,
IL 60510\label{FNAL}}

\altaffiltext{2}{Dept. of Physics, Applied Physics and Astronomy, Rensselaer
Polytechnic Institute Troy, NY 12180\label{RPI}}

\altaffiltext{3}{
Department of Astronomy,
Ohio State University, 140 West 18th Avenue, Columbus, OH 43210.
\label{OSU}
}

\altaffiltext{4}{Dept. of Physics \& Astronomy, CSCE: Center for the Study of Cosmic Evolution, and JINA: Joint Institute for Nuclear Astrophysics, Michigan State University, E. Lansing, MI  48824\label{MSU}}

\altaffiltext{5}{Apache Point Observatory, P.O. Box 59, Sunspot, NM 88349.
\label{APO}}

\altaffiltext{6}{Department of Physics, University of Ljubljana, 
Jadranska 19, 1000 Ljubljana, Slovenia e-mail:
paola.refiorentin@fmf.uni-lj.si\label{Slovenia}}

\altaffiltext{7}{Max-Planck-Institut f\"{u}r Astronomie, K\"{o}nigstuhl 17, D-69117 Heidelberg, Germany\label{MPH}}

\altaffiltext{8}{
Department of Astronomy, Case Western Reserve University, Cleveland, OH 44106.
\label{CASE}
}

\shorttitle{Sgr Tidal Tails}

\shortauthors{Yanny, Newberg et al.}

\begin{abstract}

We show that the Sagittarius dwarf tidal stream can be traced with
very red K/M-giant stars, selected from
SDSS photometry.  A subset of these stars are spectroscopically
confirmed with SEGUE and SDSS spectra, and the distance scale
of 2MASS and SDSS M giants is calibrated to the RR Lyrae
distance scale.  The absolute magnitude of the
K/M-giant stars at the tip of the giant branch is $M_{g_0}=-1.0$.  
The line-of-sight velocities of the M giant and BHB stars that are
spatially coincident with the Sgr dwarf tidal stream
are consistent with those of previous authors, reinforcing the need for new
models that can explain all of the Sgr tidal debris stream observations.
We estimate stellar densities along the tidal tails that can be used to help
constrain future models.  The K/M-giant, BHB, and F-turnoff stars in the
lower surface brightness tidal stream that is adjacent to the main leading 
Sgr dwarf tidal tail have velocities and metallicities that are
similar to those of the stars in the leading tidal tail.  The ratio of
K/M giants to BHBs and BHBs to F-turnoff stars are also similar for both
branches of the leading tidal tail.
We show that there is an additional low-metallicity tidal stream near
the Sgr trailing tidal tail.

\end{abstract}


\section{Introduction}

Our current model of galaxy formation includes the hierarchical
merging of galaxies, first proposed by \citet{1978ApJ...225..357S}.
Evidence for this includes the observation of dwarf galaxies and
globular clusters that are even now going through the process of tidal
disruption that will eventually assimilate stars from what were once
distinct gravitationally bound objects into the Milky Way spheroid.
The Sloan Digital Sky Survey (SDSS; Abazajian et al. 2009) 
and the follow-up SEGUE survey
\citep{yetal09} have provided astronomers with a detailed look
at tidal disruption of dwarf galaxies and globular clusters in
the spheroid of the Milky Way
\citep{ynetal00,ietal00,netal02,yetal03,2006ApJ...639L..17G,
2006ApJ...637L..29B, 2006ApJ...643L..17G, betal06a, 2006ApJ...645L..37G,
2006ApJ...651L..29G,2007ApJ...657L..89B,netal07,
2008arXiv0811.3965G,2009arXiv0901.4046W}.  

The first significant tidal stream to be discovered,
and also the most prominent halo stream, is from
the disintegrating Sagittarius dwarf spheroidal galaxy \citep{ietal94,ietal01}.
The Sagittarius dwarf tidal tails have been traced nearly completely around 
our Galaxy using 2MASS M-giant stars \citep{mswo03}; it is inclined
only $13.5^\circ$ from a polar orbit.  Since the shape of the
dark-matter potential determines the precession of the tidal-debris orbits,
and the lumpiness of the dark matter could scatter stars out of debris
streams, the Sgr dwarf tidal debris stream may be an excellent laboratory 
with which to study the distribution of (presumed) dark matter in and 
around the Milky Way.
 
The radial velocities of the M giants in the Sgr tidal stream, however,
have presented a puzzle for understanding our Galaxy's
gravitational potential.  Since Sgr tidal debris is fairly well constrained
to a plane, \citet{ietal01} and \citet{mswo03} have argued that the potential
must be close to spherical.  On the other hand, \citet{h04} argues that
the debris could be close to the orbital plane in a prolate potential if
it was stripped within the last 1.6 Gyr, and that the velocities of
the leading tidal tail favor a prolate dark-matter potential.  This is further supported by data on RR Lyrae stars from \citet{vzg05}. Other
studies have identified stars that are apparently
in the trailing tail 80 kpc from the Sun \citep{netal03}, and are near
the Sgr orbital plane.  This suggests that the potential is nearly spherical
or slightly oblate.  The leading tidal tail appears to be bifurcated
\citep{betal06a}.  Detailed modeling of the Sgr leading 
tail(s), making the assumption that the smaller of the two streams in
the North Galactic Cap is an extension of the trailing tidal tail, and
is thus a full orbit behind the main leading tidal tail, 
confirm this spherical dark halo finding \citep{fetal06}.
The tilt of the tidal tails with respect to the Sgr orbital
plane and other velocities of 2MASS M giants in the Sgr stream, however,
suggest that the potential is oblate \citep{ljm05}.
This puzzle has so far prevented measurement of the even low-order
properties of the halo potential.

More data and better models are clearly required. Both photometric 
imaging, which provides coordinates, estimated distances (from 
photometric parallax), population information (from colors), and 
stream stellar densities as a function of position along the stream, 
as well as spectroscopic information, including radial velocity, kinematics, and 
metallicity and gravity estimates, are needed for a large set of Sgr 
stream stars to uniquely constrain an orbit within a potential which 
may change as a function of radial distance from the Galactic 
center \citep{allgood06}.

\citet{netal07} suggested that there was sufficient confusion in the stellar
membership of the stars in the Sgr leading tidal tail that it was possible
that the velocities of the leading tidal stars were incorrect.  Figure
1 reproduces the stellar density polar plot for photometrically selected
SDSS F-turnoff stars ($0.2<(g-r)_0<0.3$, $(u-g)_0>0.4$, $21<g_0<22$) from the
lower panel of Figure 6 in \citet{netal07}, but here we also show the
star density in the South Galactic Cap.  Using a typical absolute magnitude
for the F stars of $M_{g_0}=4.2$, the Sgr stream is traced for distances 
of 23 to 36 kpc from the Sun.  The Sgr stream disappears on the right side
of the top panel where it is further from the Sun, and is faint on the
left side of the top panel where it is closer to the Sun.
Notice in Figure 1 that there is a clear bifurcation of the Sgr dwarf tidal
stream as traced by F-turnoff stars (noted by previous 
authors) above and below $(l,b) = (255^\circ,75^\circ)$.  

In this paper we isolate stars that are spatially associated with the Sgr
leading and trailing tidal tails, and so ensure that the velocities we are
measuring are in fact from stars that are part of the tidal debris
streams.  Using these spatially selected
stars, we determine that the velocities from the \citet{ljm05} paper
are correct, so the puzzle regarding the halo shape still remains.

We show that the velocities, metallicities, and stellar populations in the
two pieces of the Sgr dwarf leading tidal tail above and below $(l,b) = (255^\circ,75^\circ)$ are indistinguishable,
which calls into question the assumption that they are different ``wraps"
of the Sagittarius stream.  \citet{netal07} showed that the smaller
of the two Sgr leading tidal tails was only slightly farther away than
the larger one.

In addition, we show in this paper that K/M giants, color selected in 
SDSS and SEGUE, can be used to trace distant pieces of the Sgr tidal stream.
\citet{mswo03} and collaborators first used this technique with the
all-sky 2MASS survey to trace the Sgr tail around the sky to
distances of $>50$ kpc.  We use SDSS and SEGUE
spectroscopy of selected K/M giants to confirm their identity as distant
giants, rather than nearby M dwarfs or other contaminants.  This
allows us to make a nearly complete fiducial color magnitude sequence
for the Sgr stream in the direction of Virgo/NGP in the SDSS $ugriz$
filter system. 

We refine the distance measurement to the Sgr structure by tying the 
K/M-giant stars to blue horizontal-branch stars (BHBs) and RR Lyraes \citep{ietal00}
previously detected in the same region of the sky.  We confirm the
result of \citet{netal07} that the leading tidal tail of Sgr stream does 
not go through the Solar position, but instead passes over the Solar
position in the North Galactic cap, and passes through the Galactic plane
in the general direction of the Galactic anticenter.

\section {Selecting K/M giants in SDSS Filters}

We show in this section that we can photometrically select the reddest
K/M0-giant stars in the Sgr dwarf spheroidal tidal stream with SDSS $ugr$
filters.  M giants are generally metal-rich stars at the tip of the
red giant branch, above the point where the horizontal branch
intersects the ascending giant branch.  They are relatively rare
in the Galaxy at high latitude $|b|>30^\circ$, and are saturated
in the SDSS ($r < 14$) if they are closer to the Sun than about
20 kpc.  
Globular clusters are generally
metal poor and therefore have more vertical giant branches in $(g-r)_0, g_0$
color magnitude diagrams.
Unfortunately, K giants are difficult to 
distinguish from K dwarfs based on SDSS 
$ugriz$ colors alone; a spectrum is required to reliably 
separate the dwarfs from the giants in the color range $0.5 < (g-r)_0 < 1.1$.

Figure 2 (upper plot) shows a $(u-g)_0$ vs. $(g-r)_0$ color-color 
diagram of 11,236 SDSS stars with $g_0 < 19$ toward 
$(l,b)= (330^\circ,61.5^\circ)$, where
the Sgr stream is 44 kpc away \citep{netal03}.  
This dataset can be reproduced by selecting from the STAR 
data table of SDSS DR7 \footnote{http://www.sdss.org} all of the objects
with $0.9 < (g-r)_0 < 1.6$, $1<(u-g)_0<4$, $200^\circ<\alpha<210^\circ$,
$-1^\circ<\delta<4^\circ$, and $g_0<19$ in the SDSS DR7 database.
The SDSS STAR table contains one instance of each point source in the photometric survey. 
We compare this with a
control field (Figure 2, lower plot) that is located at at the same Galactic
latitude but at the mirror image position on the opposite side of the
Galactic center, $(l,b)=(30^\circ, 61.5^\circ)$, so the number of stars is
expected to be similar (except that the Sgr dwarf stream does not
pass through that position).  The mirror field contains 10,485 stars
selected with the same criteria as the Sgr stream field, except with
$219^\circ<\alpha<229^\circ$, $20^\circ<\delta<25^\circ$.
This diagram clearly shows an excess of
red stars below the main sequence in the direction of the Sgr tidal
stream.  An astute reader will notice that the M-dwarf main-sequence stars
are somewhat redder in the control field than they are in the direction
of the Sgr stream.  That is because all of the stars have been
dereddened with the full reddening correction from \citet{sfd98}; since
M dwarfs are in front of most of the dust, they have been 
over-corrected.  In the direction of $(l, b)=(330^\circ, 61.5^\circ)$, 
E(B-V)$=0.028$, and in the direction of $(l, b)=(30^\circ, 61.5^\circ)$, 
E(B-V)$=0.044$.  Therefore, the M dwarf stars have been over-corrected
by a larger amount in the control field, explaining their apparent
redder color.  The giants, on the other hand, are all at distances well beyond the dust.

In Figure 2 we show two
regions of color-color space in which Sgr K/M giants are separated from
disk main-sequence stars.  The long truncated ``diagonal" strip with $$(u-g)_0 >3, 1.1 < (g-r)_0 < 1.6, 1.5 (g-r)_0 + 1.09 < (u-g)_0 < 1.5 (g-r)_0 + 1.59\eqno(1)$$
is most useful for detecting K/M giants at high Galactic
latitude.  In the SDSS, the photometric errors on $u$-band become $> 0.3$ mag at
about $u \sim 22$, corresponding to $g \sim 18.5$  
for these red stars, which means we can confidently photometrically distinguish the reddest K/M
giants from dwarfs out to distances of about 80 kpc from the Sun.  At lower Galactic latitudes,
there are so many dwarf M stars that even the small fraction that fall
at large distances from the locus of dwarf stars make it impossible to
distinguish the K/M-giant stars along this whole strip.  However, the very
reddest stars in the color-color box $1.35 < (g-r)_0 < 1.6$, $3.4 <
(u-g)_0 < 4.0$, shown as a rectangle in the lower right of Figure 2, are well separated and can still be used to trace
Galactic structure.  We explored the possibility of including $r-i$
and $i-z$ information to achieve better separation of the K/M dwarfs and
giants, but only tiny improvements to the selection efficiency were
obtained.

\section{Properties of Sgr K/M giants}

Using the color selection technique from the previous section, we are
able to select K/M-giant stars in the Sgr dwarf tidal stream, and place
them on a Hess diagram similar to that of Figure 5 in \citet{netal02}.  In that
diagram it is not possible to distinguish the M giants due to the more
numerous foreground of M-dwarf stars.  We selected stars that were on
the Sagittarius stream from the same part of the sky that were used in
the 2002 version of the figure.  One slight difference is that, when
the 2002 paper was written, the database was not fully developed, so we
used only stars in particular stripes; here we select 
photometry from objects classified as STAR in SDSS DR7.  This
slightly changes the epoch of the imaging data used near the stripe
boundaries.  The Sgr stars are selected with
$200^\circ<\alpha<225^\circ$, $-1.25^\circ < \delta < 3.75^\circ$, $14<g_0<19.5$,
$1.16<(g-r)_0<2.5$, and $ugr$ colors within the diagonal K/M-giant strip
cut of Figure 2 (equation 1).  Because there are very few 
spheroid K/M giant stars outside of the Sgr dwarf tidal stream, 
it was not necessary  to perform a
background subtraction of stars from regions of the sky that do not 
contain the Sgr dwarf galaxy.

In Figure 3a, we show the Hess color magnitude diagram for the stars
with colors of K/M giants (amplified by 10x) superimposed on the Hess
diagram of DR7 stars in the box  $200^\circ < \alpha < 225^\circ, -1.25^\circ < \delta < 3.75^\circ$.  The gap in K/M stars between $g_0 = 18$ and 
$g_0 = 19$ supports our contention that the population of photometrically identifed 
K/M-giant candidates is distinct from that of disk M dwarfs.
Figure 3b shows
fiducial loci from \citet{anetal08} and \citet{clemetal08} for the clusters M3 with [Fe/H] = $-1.5$ (cyan), M92 with [Fe/H] =$-2.1$ (blue), and
M71 with [Fe/H] $\sim -0.8$ (red) and shows 
the spectroscopically confirmed K/M giants in the region
of the Sgr Northern leading tail (marked with plus signs and crosses,
see section below).

From the relative apparent magnitudes of the K/M-giant stars and the
A-type stars in the Sgr dwarf tidal stream, we can calculate
the absolute magnitude of these K/M-giant stars.  Our previous work
\citep{netal03} shows that the Sagittarius BHB stars 
in stripes 10 and 11 are at $g_0=19.0$, consistent with Figure 3 here.  
The tip of the red giant branch 
is at $g_0=17.3$ for the stars in this direction, again derived from
averaging the stars in Figure 3.  If the horizontal-branch
stars have an absolute magnitude of $M_{g_0}=0.7$, then the tip of the
M-giant branch in the Sgr dwarf tidal stream is at $M_{g_0}=-1.0$, 
and $M_{r_0}\sim -2.3$, estimated from the typical M-giant color of
$(g-r)_0 = 1.3$.
The estimate that BHB stars in our color range have $M_{g_0}=0.7$ was
obtained in \citet{ynetal00}, by converting the absolute magnitude for halo
RR Lyrae stars from \citet{letal96} into the $g$ filter, and noting
that the BHB stars are about the same absolute magnitude as the RR
Lyrae stars.  Since this determination rests on several approximate
steps our distance scale is not absolute, however, it enables measurement of the
relative magnitudes of the different types of stars that have been
used as distance indicators in the Sgr stream.

Note here that \citet{chouetal07} measured a metallicity gradient along the
tails of the Sgr dwarf tidal tails that could affect the average absolute
magnitude of K/M giants as a function of position on the stream.  This gradient is in
addition to the scatter in absolute magnitude for these stars that
arises due to surface gravity and metallicity differences between stars
in the same stream position.  We attempted to confirm the metallicity
gradient by measuring the difference in apparent magnitude between the
BHB stars and the K/M-giant stars as a function of position along the stream,
but were unable to form any firm conclusions because the stream is not well-
separated from foreground stars by photometry for most of the length of the 
stream (see Figures 6 and 8), and the vast majority of the spectra are within 
a $30^\circ$ piece of the stream at $270^\circ < \Lambda_\odot < 300^\circ$ 
(see section 8).  The difference in apparent magnitude between the BHB and
K/M-giant stars was calculated for Sgr stream stars with $240^\circ 
< \Lambda_\odot < 270^\circ$ and with $270^\circ < \Lambda_\odot < 300^\circ$.
This apparent magnitude difference was about 1.7 magnitudes for $240^\circ 
< \Lambda_\odot < 270^\circ$ and 1.6 magnitudes for 
$270^\circ < \Lambda_\odot < 300^\circ$, but consistent with no metallicity
change over this section of the tidal stream.  It should be noted that
we are using primarily photometrically selected M-giant candidates in
this study, so the results are likely to be affected by the presence of
many K giants in the tidal debris.  This study needs to be repeated with
many more spectra of both BHB and M giants.

Next we attempt to fill in the Sgr giant branch in the
$0.7<(g-r)_0<1.1$ region by selecting spectra of stars with radial
velocities that are the same as other stars in the Sgr stream.  We
selected Sgr giant branch candidates from all of the spectra in SDSS
DR7 with $200^\circ<\alpha<225^\circ$, $-1.25^\circ<\delta<3.75$,
log $g < 4.0$, $0.4<(g-r)_0<1.6$, $g_0<19.5$, and $26 < v_r < 56~\rm  km
~s^{-1}$.  This selection netted 51 candidate K/M-giant spectra, which we
then examined individually by eye.  Thirty-three of the stars we
classified as K/M giants; the other stars were G stars and M dwarfs.
Note that this exercise was done to determine the colors and magnitudes
of K and M giants the the Sgr dwarf tidal stream, and not to assess the general success of the
color selection in identifying giant-branch stars.  Of the 51 candidate
K/M-giant stars, 25 are in the $ugr$ diagonal K/M-giant color selection box.
Of those twenty-five, one is an M dwarf, 9 are K giants, and 15 are M giants,
but recall that these stars were also pre-selected on surface gravity and
velocity.  A better way to assess the contamination of the Sgr K/M-giant
sample is to compare the density of stars in the stream to the density
of stars in the field in Figures 2 and 6.  In Figure 2, there are 94 K/M-giant
star candidates in the Sgr dwarf tidal stream and 30 in the off-stream field,
suggesting that 68\% of the selected stars in this region are actually members
of the Sgr stream.  Of these, about 60\% are expected to be M giants rather
than K giants.  In Figure 6, one can see that the background contamination
increases close to the Galactic plane, so the fraction of stars that are
in the stream will be lower.

In Figure 4 we show two selected Sgr K/M-giant tidal stream spectra, plus, for
comparison, one spectrum that we classified as an M dwarf.  The
stars were classified as M dwarfs if they had a strong and narrow
Na absorption line, the Mg triplet at 5200\AA\ was
prominent, and there was strong Ca at 4226\AA; otherwise we classified 
it as a giant.  We saw two distinctly different types of giant spectra,
as can be seen by comparing the top two spectra in Figure 4.
Stars with strong TiO bands, which we identify as M giants, as 
seen in the second spectrum in Figure 4, are shown as 
plus signs in Figure 3b.  Stars with spectra more like the top spectrum 
in Figure 4 tended to be bluer, and are K giants, as 
represented by the crosses in Figure 3b.

The giant stars in the Sgr stream were plotted on the Hess diagram in
Figure 3b, allowing us to better trace the Sgr giant
branch.  The giant branch, which previously couldn't
be seen redder than $(g-r)_0 >0.6$, is now completed from $(g-r)_0 =0.75 $
to $1.6$.  We verified that the color-magnitude diagram for the stars
that we spectroscopically identified as G stars and M dwarfs does not
show a coherent giant branch, validating our ability to correctly classify the
SDSS spectra.

Comparison of the fiducial SDSS photometric sequences of known
globular clusters \citep{anetal08} indicates that the BHB stars in the
Sgr tidal stream have metallicities of $-2.3 < \rm [Fe/H] < -1.6$.  RR Lyraes in the Northern leading tail have metallicities of $\rm [Fe/H] = -1.76\pm 0.22$ as shown by \citet{vzg05}.  The  Stellar Structure Parameters Pipeline
(SSPP) \citep{letal08a,letal08b,apetal08} generates metallicities for 
all stars in SDSS + SEGUE (over the color range $0.3 \le (g-r)_0 \le 1.3$, and with spectra
of sufficient signal-to-noise); the 33 Sgr K/M-giant stars have [Fe/H] $\sim -0.8\pm 0.3$.  We note that the absolute metallicity determinations of the SSPP are not 
well-calibrated for M giants.  However, \citet{chouetal07} have obtained high-resolution metallicities of some Sgr stream M giants in this part of the sky, and
find metallicities consistent with these values.


\section{Comparison of Sgr K/M giants in SDSS and 2MASS}

We obtained infrared photometry from the 2MASS catalog \citep{skrutskie06} 
for each of the Sgr K/M-giant stars selected in the previous section, 
using the matching program from the US National Virtual Observatory.   
Figure 5 shows stars (J-K$_S$)$_0$, (J-H)$_0$ colors of the K/M giant 
candidates selected from $ugr$ photometry, both for the strip cut and the 
very red box that can be used at low Galactic latitude.  The 2MASS colors
have been dereddened using the procedures described in \citet{mswo03}.  In the 
same figure we show JHK$_S$ photometry for spectroscopically confirmed 
Sgr K and M giants, and the JHK$_S$ selection box used by \citet{mswo03} to 
select K and M giants from Sgr over the entire sky.  

Examining Figure 5 in detail, we note that the 2MASS $(J-H)_0$ color is 
good at separating dwarfs with $0.4 < (J-H)_0 < 0.7$ from giants 
with $0.7 \le (J-H)_0 < 0.95$, while $(J-K)_0$  can be additionally used
to somewhat separate K giants with $0.8 < (J-K)_0 < 1.0$ from M giants with $1.0 < (J-K)_0 < 1.2$.  

We select similar K- and M-giant candidates to the \citet{mswo03} selection, but the
population of K and M giants selected optically in $ugr$ by the diagonal strip cut of Eq. 1 includes 
some bluer stars in $J-H, J-K$, while the very red SDSS $ugr$ selection box of Figure 2 excludes most of the 
stars on the blue side (in $J-K$) of the \citet{mswo03} selection box.


\section{Tracing Sgr in K/M Giants}

Now we select K/M-giant candidates via the strip color cut (Eq 1)
over the whole of the SDSS+SEGUE imaging area.  We used only regions of the
sky with low reddening ($E(B-V)<0.25$).  In order to eliminate some
of the nearby disk M dwarfs from the sample, we eliminated stars from the sample
if they had proper motions of more than 6 mas $\rm {\rm yr}^{-1}$ (note that $|\mu_b|>6$ and
$|\mu_l|>6$ mas yr$^{-1}$ is required if a proper motion is measured at all in the
SDSS DR7 proper motions table, where the typical proper motion error 
for a star is typically 3 mas $\rm yr^{-1}$).  Using an estimated distance to 
each giant candidate assuming the absolute
magnitude is $M_{g_0}=-1$, as determined in \S 3, and also using the
definition of the Sgr dwarf coordinate system as defined by
\citet{mswo03}, we convert $(l,b)$ and distance to $(X_{Sgr}, Y_{Sgr}, Z_{Sgr})$
for all K/M-candidate giant stars.  In this coordinate system, the axes $X_{Sgr}$ and $Y_{Sgr}$ are
in the Sgr dwarf spheroidal orbital plane, and are not far from the
Galactocentric $X$ and $-Z$ coordinates, respectively.  The $Z_{Sgr}$ coordinate direction is
perpendicular to the Sgr dwarf orbital plane, and is approximately in the
Galactocentric $+Y$ direction.

Figure 6a shows 5,402 stars within 5 kpc of the Sgr dwarf orbital plane, 
$|Z_{Sgr}| < 5$ kpc.  The leading Sgr tail arcs 
over the Sun's position and intersects the plane of the Milky Way at 
approximately $X_{Sgr}= -25$ kpc (the Sun is at -8 kpc, so that is about 17
kpc from the Sun).  Additionally, candidate Sgr K/M giants south of the plane are clearly 
present.  The positions of these
stars are consistent with the positions of BHB stars in Figure 1 of
\citet{netal07}.

Our next step is to spatially isolate those stars that are in the Sgr
stream in three dimensions.  
Part of the motivation for this is to separate the Sgr members from 
the closer Virgo stars, a very
confusing region of the halo with likely several structures present
\citep{vetal01,jetal06,netal07,kdp09}.

We draw two parabolic arcs to define the limits of the leading tidal tail:
$Y_{Sgr}<0$, $0.025(X_{Sgr}-11)^2-53 < Y_{Sgr}$, and 
$Y_{Sgr} < 0.028(X_{Sgr}-8)^2-32$, where all distances are in 
kpc.  In the south Galactic cap we selected stars in the trailing tidal
between the parabolas defined by:
$Y_{Sgr}>0$, $-0.16(X_{Sgr}+10)^2+16>Y_{Sgr}$, and 
$Y_{Sgr}<-0.010(X_{Sgr}+18)^2+35.$
Figure 6b shows in red the 472 candidate K/M-giant stars that were selected in the region
of the leading tidal tail, and the 133 candidate K/M-giant stars that were selected
in the region of the trailing tidal tail.

Figure 7 shows the $(l,b)$ positions
of the final set of candidate K/M-giant stars that are spatially coincident with the Sgr
dwarf tidal stream defined by the annular cuts above.  The upper panel shows positive $b$ and the lower 
panel shows negative $b$ (below the Galactic plane).  The light gray 
shading represents the sky covered by the SDSS and SEGUE imaging 
surveys.  The darker black regions represent areas where the Sgr tidal 
stream stars are present.  
Since the stars were selected to be at the same distance
as the Sgr stream, and within 5 kpc of the Sgr orbital plane, the footprint
of the stream on the sky is narrower on the right side of the figure where the
stream is 50 kpc away, and wider on the left side where the stream is only
20 kpc away from the Sun.  We do not see a bifurcation of the Sgr stream
in this figure, though this can be easily accounted for by the low numbers of
of K/M-giant candidates.


\section{Sgr Traced in A Stars}

We now do the same analysis as performed for the K/M stars, except we
choose photometrically selected BHB stars, which are also very prominent in the Sgr tidal 
streams.  The A-type stars were selected from the SDSS DR7 STAR database, 
with
$15<g_0<21$, $-0.3 < (g-r)_0 < 0.0$, and $0.8<(u-g)_0<1.5$.  An additional
color cut in $ugr$ was applied that separates these blue stars into those
that are more likely to be high surface-gravity stars and those that are
likely to be low surface-gravity BHB stars \citep{ynetal00,netal07}.  
Just as in the K/M-giant candidate selection, we eliminated stars with proper motions
larger than 6 mas yr$^{-1}$ and in areas of the sky with $E(B-V)<0.25$.
The positions of the 32,881 BHB stars within 5 kpc of the Sgr 
dwarf tidal stream are shown in Figure 8a.

Because the K/M-giant candidate selection is cleaner (the photometrically
selected BHBs have significant contamination from blue-straggler stars
and BHBs in the spheroid), the spatial selection for leading and trailing tail BHB stars was
chosen to be identical, assuming the candidate K/M giants are at $M_{g_0}=-1.0$
and the BHBs are at $M_{g_0}=0.7$.  
Figure 8b shows in red the positions of all the
selected BHBs that are in the same areas of sky as the candidate K/M giants in Figure 6b,
but with $g_0$ limits that are shifted by 1.7 magnitudes.

We note several interesting features in Figure 8 in the region beyond 40 kpc from
the Sun.  There is a significant group of BHBs at about 80 kpc from the Sun,
near $(X_{Sgr},Y_{Sgr}) = (-80,-40)$ kpc, which are believed to be associated with the Sgr stream,
and have been discussed in detail by \citet{netal03}.   Additionally, in their study of
the Sagittarius stream, \citet{betal06a} note distant ($d\sim 50 \rm ~kpc$) structures manifesting
themselves as double sub-giant branches in color-magnitude diagrams along the stream.
The fainter turnoffs which they note at $180^\circ < \alpha < 190 ^\circ$ correspond roughly to
the overdensity of BHB points at $(X_{Sgr},Y_{Sgr}) = (-10, -50)$ kpc, behind the main upper branch
of the Sgr stream at $(X_{Sgr},Y_{Sgr}) = (-10, -30)$ kpc.

Figure 9 shows the $(l,b)$ polar plots of BHB stars selected in Figure
8b that are within 5 kpc of the Sgr dwarf orbital plane.  The density pattern
is very similar to the density pattern for candidate K/M giants shown in Figure 7.
The bifurcation is not apparent, probably because the star counts are not
high enough to see the two streams clearly.
The dark spot on the left side of the top
panel in Figure 9 near $(l,b) = (185^\circ,25^\circ)$ is a lower latitude
region where the Monoceros stream is also present at these distances ($d\sim 20  \rm ~ kpc$) from the Galactic plane.  Thus we are not certain of the
association of this BHB overdensity with the Sgr stream.

\section{Sgr Traced in F Stars}

The F-turnoff stars are also selected for comparative analysis.  
They are more numerous, but we begin to lose sensitivity to these intrinsically fainter
 stars in the SDSS at distances beyond 40 kpc from the Sun, 
so we do not sample the most distant parts of the stream in the 
North Galactic Cap.  We photometrically selected F-turnoff stars from the SDSS DR7
STAR database with $0.1<(g-r)_0<0.3$, $19 <g_0<24$, $-6<\mu_l<6$, $-6<\mu_b<6$ mas yr$^{-1}$,
$E(B-V)<0.25$, and $-5 < Z_{Sgr}<5$ kpc, where the conversion from $(l,b,g_0)$ to 
$(X_{Sgr},Y_{Sgr},Z_{Sgr})$ assumes $M_{g_0}=4.2$.  We also eliminated stars near
the globular clusters M~53 and NGC~5053 by eliminating stars in the region
$331^\circ<l<337^\circ$ and $79^\circ<b<81^\circ$.  This results in a sample of
710,832 stars.  Figure 10 shows the density of F-turnoff stars in the Sagittarius
orbital plane.  The Sgr stream artificially appears much broader in F-turnoff stars, 
since these stars have a wider range of absolute magnitudes; when we calculate
each star's position we assume a single, average, absolute magnitude so there are
large distance errors in the positions of these stars in Figure 10.  In the upper right
corner of this diagram we lose most of the stars in the Sgr dwarf tidal stream
because they are too distant to be seen in SDSS data.  The background counts of
F-turnoff stars are very high near the Sun, and those with $g_0 < 19$ are 
removed from the figure.

Nevertheless, we select F-turnoff stars $|Z_{Sgr}| < 5 $ kpc as we 
selected BHB and candidate K/M-giant stars in the previous two sections.  
Because there are much larger distance errors, we select a somewhat smaller fraction of the
Sgr F-turnoff stars this way.  The 223,977 stars that are positionally
coincident with the Sgr stream are shown in Figure 11.  Even though we lose
the Sgr stream at large distances, on the right side of the upper panel, we do not
lose them as quickly as for the magnitude-limited sample in the upper panel of
Figure 1.  The larger number of F-turnoff stars allows us to see the
bifurcated leading tidal tail above and below $(l,b) = (255^\circ,75^\circ)$.  
Some SDSS striping, an artifact of the SDSS's photometric calibration near 
its magnitude limit $g \sim 23.5$, is seen in the density distribution of the upper
 panel of Figure 11.  

\section{Velocities along the Sgr Stream}

We now select stars that are positionally coincident with the Sgr leading
and trailing tidal tails, and plot their line-of-sight, Galactic standard
of rest velocity, $v_{gsr} =  {\rm rv}+10.1~ cos~b~cos~l+ 224.0~cos~b~sin~l+ 6.7~sin~b ~\rm km~s^{-1}$, where rv is the heliocentric radial velocity, 
as a function of position along the stream, $\Lambda_\odot$.
We selected from SDSS DR7 all of the SDSS/SEGUE velocities for K/M-giant 
stars in Figure 7 for which spectra exist.  Neither SDSS nor SEGUE targeted
giant stars this red; most of the 55 selected K/M-giant spectra are from a small set of
special plates in a very restricted part of the sky.  Since we have spectrally
determined surface-gravity information for horizontal-branch (HB) stars
that will allow us to distinguish them from the much more numerous
F dwarfs, we can select HB-star spectra with a much broader color range  
that we used to photometrically select the BHB stars in Figure 9.  
The spectroscopically selected BHB-star sample includes all SDSS DR7 spectra
of point sources with $-0.3<(g-r)_0<0.35$, $0.8<(u-g)_0<1.5$, $1.0< \rm log~ g< 3.75$,
and $-5 < Z_{Sgr}<5$.  The annular cuts of \S 5 are performed. 
Using these criteria, 999 HB stars were selected.

We show in the upper panel 
of Figure 12 a plot similar to that of \citet{ljm05} Figure 12, showing 
the Galactocentric, line-of-sight velocity of all K/M giants with radial velocity information, 
as well as the 999 HB stars with velocities, as a function of angle along the Sgr tidal 
stream.  The heliocentric angle along 
the stream, $\Lambda_{\odot}$, is the same as used above.

When we initially made Figure 12, we noticed that the velocities of the 
HB stars near $\Lambda_\odot=100^\circ$ did not match the velocities
measured in the \citet{ljm05} paper.  We looked at the metallicities
of the BHB stars and determined that in the South Galactic Cap,
near $\Lambda_\odot=100^\circ$, there was a cluster of BHBs with lower
metallicities than the other Sgr BHB stars, and higher $v_{gsr}$ than
the other Sgr K/M giants and BHBs.  Therefore,
we split the BHBs into higher metallicity ($\rm [Fe/H]_{\rm WBG} > -1.9$) and lower 
metallicity $\rm [Fe/H]_{\rm WBG} < -1.9$ subsets.  The WBG subscript refers to
the technique of estimating metallicity for a blue star based on its estimated
$\rm T_{\rm eff}$ and Ca K line strength of \citet{wbg99}. 
The SDSS SSPP measures metallicities 
by several techniques, and includes an `FeHa' ([Fe/H] adopted) average metallicity
that is generally accepted as the best overall measure of metallicity.  
However, we have found that the WBG metallicity, which was developed specifically 
for BHB stars, is a better measure for these blue stellar 
types, so we use that measure for HB stars throughout this paper.

The K/M giants and the higher-metallicity
BHB stars trace the Sgr dwarf tidal stream, and are a good match to the
velocities determined by \citet{ljm05}.  The low-metallicity BHB stars,
while they are at a similar apparent distance as the Sgr dwarf tidal stream, 
have distinct velocities in the trailing tidal tail (upper panel of Figure 12).
In the North Galactic Cap, there are many high- and low-metallicity BHB
stars with a broad range of velocities, as one would expect in the spheroid,
the K/M giants and some of the higher-metallicity BHB stars follow the
Sgr stream, and there is a cluster of stars with zero or slightly
negative $v_{gsr}$ at $\Lambda_\odot=190^\circ$ (which are presumably
associated with the Monoceros stream complex, or a distorted disk, that
is found at low Galactic latitude near the anticenter).

Spectra of stars in the field associated with the globular 
clusters NGC~5053 and M~53, which overlaps the Sgr north high-latitude branch,
are explicitly removed from this study.

In the Sgr stream we find both low- and high-metallicity BHB stars.
Note the very low-metallicity BHBs in the Virgo Overdensity at 
$\Lambda_\odot=230^\circ$, $v_{gsr}=130$ km s$^{-1}$ in the upper panel of
Figure 12.  
Additionally, a close examination of spectra associated with 
density enhancements in the lower panel of Figure 9, where the Sgr 
stream crosses near the South Galactic Pole, also shows a second stream, 
apparently independent of Sagittarius, and not in the Sgr plane, 
of very low-metallicity stars ([Fe/H] $\sim -2$).  These low-metallicity 
stars will be explored in a 
forthcoming paper \citep{cetuspolarstream}. 

Figure 13 highlights the metallicity difference used to help identify
this apparent additional population, distinct from Sgr in the south.  
Figure 13(a) contains two outlined boxes. 
The rectangular box has 22 high-metallicity BHBs (black dots) and 41 low-metallicity BHBs (blue crosses),
a ratio of 1:2 high:low.  The diagonal Sgr box, however, has 60 high-metallicity
BHBs and 37 low-metallicity stars, for a reverse ratio approaching 2:1. The
distances to the objects are shown in Figure 13(b),  where the objects in
the rectangular (non-Sgr) box are presented as (green) triangles (low metallicity)
and squares (high metallicity).  Here too, the distribution of the low-metallicity BHBs in
the rectangular at 30 kpc appears distinct from that 
of the higher-metallicity stars in Sgr, which slopes from d=20 kpc
to d=30 kpc as $\Lambda_\odot$ increases from $75^\circ$ to $120^\circ$.

The Sgr BHB stars and the K/M-giant stars 
with SDSS and SEGUE spectra trace the Sgr dwarf tidal stream with 
essentially the same velocities as those 
measured by \citet{ljm05}, using stars that follow the density
enhancement of the the Sgr stream.  The conjecture put forth by
\citet{netal07} that the Sgr stream velocities in \citet{ljm05} might 
have been measured for stars at the wrong distance appears to be
incorrect.  Therefore, it remains a challenge to 
fit a model for Sgr dwarf tidal disruption to the measured 
velocities in the leading tidal tail.

\section{Analysis of the ``Bifurcated'' leading tidal tail}

In this section we compare spectra of the stars in the two tidal 
debris streams identified by \citet{betal06a}, and apparent in the top panel of
Figure 1.  In the North Galactic Cap, the two pieces of the stream can be
separated in $Z_{Sgr}$.  Figure 14 (upper panel) shows the positions of all of the SDSS
and SEGUE spectroscopically selected HB stars, as selected in the previous 
section, (and shown in Figure 12) that have $-5<Z_{Sgr}<5$ kpc.  The smaller debris stream 
at higher Galactic latitude, $Z_{Sgr}>0.04 X_{Sgr} + 1.0$, is shown as cyan
dots in the figure.  The larger debris stream, $Z_{Sgr}<0.04 X_{Sgr} + 1.0$,
is shown as magenta dots.  Although the larger stream at lower Galactic latitude
has more stars, and presumably more HB stars, it is more sparsely sampled.

Figure 15 shows all of the stars in Figure 12 that are in the higher Galactic
latitude piece of the leading tidal tail, and Figure 16 shows all of the stars in
Figure 12 that are in the lower Galactic latitude piece of the leading tidal tail.

The K/M giants in both the Northern and Southern branches have velocities that are
consistent with the full set of spectra.  The difference in the $\Lambda_\odot$
positions of the concentration of K/M giants in each branch is a selection effect
in the data and not due to any intrinsic difference between the streams.  In the
lower panels of Figure 15 and 16, we see that the K/M giants in the higher Galactic latitude stream are on
the closer side of the selection box, while the K/M giants in the lower Galactic 
latitude stream are on the further side of the selection box.  In contrast,
\citet{netal07} found that the northern branch was farther away at 
$\Lambda_\odot \sim 220^\circ$.  This suggests that the two pieces of the stream
cross each other in distance.

There is also no difference in the velocities of the Sgr BHB stars as a function of 
$\Lambda_\odot$.  Since the more northern branch of the leading tidal tail is
extremely low surface brightness in the region $180^\circ<\Lambda_\odot<240^\circ$,
the Sgr stream is not apparent in Figure 15 in this region.  

In the lower panel of Figure 14 we show all of the spectra from the upper panel
that also have the velocities of the Sgr dwarf leading tidal tail.  These were
selected as stars that have 
$0.10(\Lambda_\odot-200^\circ)^2-150<v_{gsr}<0.17(\Lambda_\odot-200^\circ)^2-90$ km s$^{-1}$,
and $180^\circ<\Lambda_\odot<300^\circ$.  This cut eliminates most of the spectra
at low Galactic latitudes near the Galactic anticenter, most of the spectra on the
Orphan Stream, and most  of the spectra that are not aligned with photometric
overdensities of F-turnoff stars in Figure 14.

\section{Metallicities of the Sgr BHB stars}

It was shown in \citet{harrigan} that, by comparing BHBs in globular 
clusters with published globular cluster metallicities, that the SSPP 
WBG metallicities for BHB stars were reasonable, but possibly a tenth of 
a dex too high.  Here, we would like
to compare the BHB stars in the two branches of the leading Sgr tidal tail,
doing an internal comparison to see if the metallicities are different.  This
check is independent of any overall offset in the metallicities determined
by SSPP, which could be a few tenths of a dex.

In Figure 17, the upper panel shows the distribution of metallicites of the HB stars in Figure
14 (lower panel).  These stars
were divided into those that have the velocities of the Sgr leading tidal tail,
as used to select the spectra in the lower panel of Figure 14, and those which
were in the upper panel but did not have velocities of the Sgr leading tidal
tail.  The Sgr BHB spectra were further divided by branch of the leading tidal
tail (cyan and magenta points in Figure 14 for the upper and lower branches respectively).  Note that both the upper and lower branches of 
the Sgr leading tidal tail stars have similar metallicity distributions.  The HB stars that do not have
the correct velocities are spread in a much broader range of metallicities,
as one would expect from a background spheroid population (black line in upper panel of Figure 17).  
The Sgr dwarf BHBs in both branches of the leading tidal stream 
have metallicities $-2.3 < {\rm[Fe/H]_{WBG}} < -1.6$.  
Since the BHBs are expected to trace only the older, more metal poor component
of the Sgr dwarf galaxy, these stars do not represent the entire range of
metallicities in the stream.  Some of these stars may in fact overlap the RR Lyrae sample studied by \citet{vzg05} in the Northern Sgr leading tidal stream.  Our determination of HB metallicity (as well as velocity) is consistent with their determination of $\rm [Fe/H] = -1.76\pm 0.22$.

The lower panel of Figure 17 justifies our definition of `higher' and `lower'
metallicities in Figure 12.  The metallicity distribution of the BHB
stars with velocities of the trailing Sgr tidal stream ($70^\circ<\Lambda_\odot<108^\circ$,
$-2.0 \Lambda_\odot+60 < v_{gsr} < -2.0 \Lambda_\odot + 120$ km s$^{-1}$) are similar to the
metallicities of both branches of the Sgr leading tidal tail.
On the other hand, stars selected with velocity ($-71 < v_{gsr} < 0$ km s$^{-1}$)
and $\Lambda_\odot$ ($85^\circ < \Lambda_\odot < 140^\circ$)
are consistent with the new debris stream discovered in the South Galactic Cap have
a lower metallicity (see Figure 13).

\citet{betal06} showed that the ratio of BHB to red-clump stars is higher in the stream vs. the
bound core of Sgr, indicating that the stream stars are from a more metal-poor
population of stars that was less tightly bound and thus was stripped first.  We studied 
the metallicity of the BHB stars as a function of $\Lambda_\odot$, and find that there
is no significant trend in BHB metallicity vs. $\Lambda_\odot$, at least along these portions 
of the tidal stream ($100^\circ < \Lambda_\odot < 115^\circ , 
~200^\circ < \Lambda_\odot < 300^\circ$).  Note that since all BHB stars have low
metallicity, it is unclear that one would expect a trend in the metallicities of BHB
stars even if there is a metallicity gradient along the stream \citep{chouetal07}.

\section{Densities along the Sgr Stream}

Figures 7, 9, and 11 graphically show the density of K/M-giant, BHB, and F-
turnoff stars, respectively, along the Sgr dwarf tidal tails.
Note that the number counts in BHB and K/M-giant stars are much higher on the
right sides of these figures in the North Galactic Cap, both because the stream 
density is higher at apogalacticon and because the stream is farther away, so 
each pixel, which represents a fixed angular area in the sky, intersects a larger
linear distance along the stream.  In this section, we construct histograms of the 
number counts of these three types of stars along the Sagittarius stream.

We plot in Figure 18 the number density of photometrically selected Sgr 
candidate K/M-giant, BHB and F-turnoff stars, selected
as a function of angle along the Sgr dwarf tidal stream, $\Lambda_\odot$, 
as defined in \citet{mswo03}.
Figure 18(a) shows the star counts from Figures 7, 9 and 11
as a function of $\Lambda_\odot$.  

In Figure 18(b), we correct the star
counts for the fraction of the width of the stream we observe, subtract the
background counts, and divide by the distance to the stream.
To estimate the fraction of the
stream width that we observe in each ten-degree slice of $\Lambda_\odot$,
we first estimate the density of objects on the sky within the SDSS footprint by using
quasar number counts as a constant density reference tracer \citep{ynetal00}.  
We select QSOs over the whole sky from the SDSS DR7
STAR data base which had $-0.05<(u-g)_0<0.4$,
$-0.1<(g-r)_0<0.3$, and $18<g_0<21$.  At each $\Lambda_\odot$ bin, we
estimated the angular cross section, $\theta$, of the stream from its distance
($\theta=$5 kpc/$d$).  In addition to the longitude-like angle $\Lambda_\odot$
along the Sgr stream, \citet{mswo03} also defined a latitude-like angle $B_\odot$
that we will use to measure the angle across the stream.  We counted
all QSOs in the $\Lambda_\odot$ range that were also within the 
$-\theta<B_\odot<\theta$ limits.  The fraction of the stream sampled by
our data is given by the number of QSOs in the $(\Lambda_\odot,B_\odot)$
bin, divided by the number of QSOs we would have expected.  The expected
number of QSOs is the typical number of QSOs per square degree, multiplied by
ten degrees of $\Lambda_\odot$, and then multiplied by $2\theta$.
The normalization factor, multiplied by 20, is indicated by the small 
circles in Figure 18.  For example, if the normalization circle is at ``20", there is no
correction factor applied.  If the circle is at ``60," then the number
counts are multiplied by three.  If the stream is not uniform density
in cross section within 5 kpc of the Sgr orbital plane, then this 
correction is not perfect.  In general, $\Lambda_\odot$ bins with data
in the center of the stream but not the edges are over-corrected, and bins
with data at the edges of the stream but not in the center are
under-corrected.

To estimate the background counts, we divided both the BHB star sample and
the K/M giant sample into distance bins that were 5 kpc deep (0-5 kpc from the
Sun, 5-10 kpc from the Sun, etc.), and then for each type of star and each
distance range we constructed binned star counts.  We estimated the background
from the portions of the histograms that were away from the Sgr stream, and
not at very low Galactic latitude.  These estimates were done only in the
North Galactic Cap where we had full sky coverage, but were applied to
both the northern and southern data sets.  To find the background in the entire
distance range of the Sgr stream selection, we numerically integrated the
star counts over each of the 5 kpc deep bins that overlapped the
area between the two parabolas that were used to select the stream stars,
for each $\Lambda_\odot$ bin.  The total background counts was then
subtracted from the corrected star counts.  If the background subtraction resulted in
negative stream density, then that bin is suppressed in Figure 18.

In a fixed angular distance along the stream, we intersect a larger linear 
distance along the stream.  Therefore, after correcting for completeness
and background we divided the counts in each bin by the distance to the
stream for each $\Lambda_\odot$ bin.  Therefore, the counts in each bin
are an estimate of the number of stars along the stream, per kpc projected
perpendicular to our line of sight.

Figure 18(b) shows that the K/M-giant and BHB star counts fall
quite rapidly from $\Lambda_\odot=300^\circ$ to $\Lambda_\odot=250^\circ$, 
as we move away from one of the stream's apogalactic orbital `turning points'
where stars are expected to pile up in density.  
The F-turnoff star number counts are low 
near $\Lambda_\odot=300^\circ$ because F-turnoff stars at the distance
to the Sgr stream there are too faint to be observed in the SDSS.  We expect
that the actual counts of F stars have similar density patterns to the
K/M-giant and BHB stars.  There are about 50\% more BHB stars than K/M giants
in the leading tidal tail, and about 300 times as many F-turnoff stars as
BHBs.  In the south, the numbers are less certain because the correction
factors are larger.  The number counts for all three types of stars are
approximately constant in the range $60^\circ<\Lambda_\odot<115^\circ$.  It
is plausible that the ratios of K/M giants to BHBs to F-turnoff stars is the
same in the trailing tail as in the leading tail.  In both the leading and
trailing tidal tails, the BHB star counts show more bin-to-bin fluctuations 
than the K/M-giant star counts. 


The last two panels of Figure 18 are similar to the second panel, except they
give star counts for the upper, c), and lower, d), branches of the leading tidal tail
only, with the split at $Z_{Sgr}=0.04 X_{Sgr} + 1.0$ kpc.  The upper leading
tidal tail has about 40\% as many stars as the lower branch, but shows the
same downward trend in star counts from $\Lambda_\odot=300^\circ$ to
$\Lambda_\odot=250^\circ$.  However, the upper-branch density falls
precipitously by $\Lambda_\odot=230^\circ$.  This is most apparent in the
F-turnoff star counts.  In the range $205^\circ < \Lambda_\odot<235^\circ$
there are 12,653 F-turnoff stars in 
the lower branch, and none in the upper branch (after background subtraction).

These density distributions and ratios of relative numbers 
of K/M giant:BHB:F-turnoff stars will be of considerable 
use in modeling the Sgr stream, since the prediction of stellar number 
density can be used as a constraint, in addition to the 
position and velocity data that has so far been used.

\section{The H-R diagram of the Sgr Stream}

The SDSS/SEGUE spectroscopy give us the opportunity to look at the H-R
Diagram of the Sgr tidal stream with much less contamination from main-
sequence stars than the photometric H-R diagram in Figure 3; the drawback
is that the spectra are not a complete sample and there are significant
selection effects as a function of color.

We selected from the SDSS DR7 SppParams database table all of the objects with spectra,
within the angular limits of the Sgr leading tidal tail in Figure 7, 9,
and 11 (those figures use limits in $Z_{Sgr}$, but since we do not know the
distance to each spectroscopic target we selected a similar area of the
sky in $\Lambda_\odot, B_\odot$).  We then selected all of the stars with
low surface gravity, $1.0< \rm log~g < 3.75$, and within the velocity range
of the leading tidal tail as a function of $\Lambda_\odot$.  This resulted
in a sample of 1,887 objects that are candidate Sgr stream stars at
a variety of distances from the Sun.

Since the Sgr stream is at different distances at different places in the
sky, we calculated a ``corrected" apparent magnitude by adding the 
quantity $-5.0*\log(d_{Sgr}(\Lambda_\odot)/ 30{\rm kpc})$ to each $g_0$
apparent magnitude.  This adjusts the apparent magnitude of each star
in the Sgr stream to the apparent magnitude it would have if it were
at 30 kpc.  

The upper panel of Figure 19 shows the adjusted apparent $g_0$ magnitude as a function of $(g-r)_0$
The imprint of the SDSS/SEGUE
selection function is discernable here (for example, SEGUE took spectra of a large number
of G dwarfs near $(g-r)_0=0.5$).  We can also clearly see the Sgr stream candidate HB stars at
a corrected $g_0$ of 18, and the very red M-giant stream candidate stars at a corrected
$g_0$ of about 16.5.

The lower panel of Figure 19 shows a color magnitude diagram for stars from Figure 12 that have velocities consistent with membership in 
the Sgr tidal stream.  The BHB stars are similar, since they were 
spectroscopically selected.  Our
K/M-giant sample extends redder than the spectroscopically selected M giants.
The SSPP does not prvide reliable surface gravities for stars redder than 
$(g-r)_0=1.3$, so they are not present in the upper panel's selected sample.

\section{Discussion and Conclusions}

From analysis of optical imaging and spectroscopy of stars in the
complex Sagittarius tidal stream, we present new results on the stellar 
populations, velocities, metallicity distributions, and spatial and density 
distributions of the tidal debris stream.  This information, combined 
with other Sgr stream data in the literature, can be used to constrain 
models for the stream orbit and thereby contribute to our knowledge of 
the shape of the Milky Way's dark-matter halo gravitational potential.  
Kinematic information is crucial to separating leading-tail candidates
from trailing-tail candidates.  Metallicity and stellar population 
information is important for distinguishing or identifying separate 
stellar populations, which may have be stripped at different times.

We conclude that:

(1) We can photometrically select the reddest K/M-giant stars in the Sgr dwarf tidal stream with
SDSS $ugr$ filters for objects down to about $g \sim 18.5$, corresponding to d= 80 kpc from the Sun.  The tip of the M-giant branch in the Sgr 
dwarf tidal stream has an absolute magnitude of $M_{g_0} = -1.0$.  The giant branch in the Sgr dwarf leading tidal tail is consistent
with those of globular clusters with [Fe/H] of $-1\pm 0.5$.
The 33 identified Sgr K/M giant stars have 
metallicities of $-0.8\pm0.2$, as measured by the SDSS DR7 SEGUE Stellar Parameters Pipeline, 
but an independent determination of these giant star metallicities should be made.  

(2) Following the technique of \citet{mswo03}, we can trace the Sgr dwarf tidal stream over the
entire SDSS+SEGUE imaging footprint.  The positions of the K/M giants are consistent with the
measured positions of Sgr BHB stars from \citet{netal07}.

(3) The line-of-sight velocities of BHB and K/M giant stars that are spatially coincident with the
Sgr leading and trailing tidal tails are the same as those measured by \citet{ljm05}.  This
indicates that the velocities are not seriously contaminated with other debris in the Virgo region, as
(incorrectly) conjectured by \citet{netal07}.

(4) We measure the density of K/M-giant stars, HB stars, and F-turnoff stars as a
function of angle along the Sagittarius dwarf tidal stream.  Within our errors,
the ratio of star counts in each of these star types is the same in both
branches of the leading tidal tail and in the trailing tidal tail.
The stellar density in the Sgr leading tidal tail decreases from the 
apogalactic turning point at $\Lambda_\odot \sim 300^\circ$ going toward the
Galactic anticenter, falling more sharply for the lower surface brightness,
higher Galactic latitude branch.
These density profiles can be used to aid in modeling of the tidal disruptions.

(5) The lower surface brightness tidal debris stream `branch' that is 
parallel to the main leading tidal tail has velocities, metallicities, and 
relative densities of K/M-giant, BHB, and F-turnoff stars that are similar 
to the main Sgr leading tidal tail.  Models that explain the upper branch 
as debris that has orbited the Galaxy at least one more time than the 
lower branch (leading tidal tail), or at least one less time (trailing 
tidal tail), must also explain the similarity in populations between these
two debris streams.  Alternatively, we suggest the two branches might be
from debris that was stripped from the Sgr dwarf galaxy at similar times.

(6) A previously unknown, low-metallicity tidal debris stream that is
spatially coincident with the Sgr South dwarf trailing 
tidal stream, but which exhibits a distinct metallicity and velocity
profile, is identified and will be the subject of a future paper.

\acknowledgments 

This research was funded by the National Science Foundation (NSF)
grant AST 06-07618.
Y.S.L. and T.C.B. acknowledge partial support from
PHY 08-22648:  Physics Frontier Center/Joint
Institute for Nuclear Astrophysics (JINA), awarded by the NSF.
P.R.F. acknowledges support through the Marie Curie Research Training
Network ELSA (European Leadership in Space Astrometry) under contract
MRTN-CT-2006-033481.
This research has made use of software provided by the 
US National Virtual Observatory, which is sponsored by the National 
Science Foundation.  This publication makes use of data products from the 
Two Micron All Sky Survey, which is a joint project of the University of 
Massachusetts and the Infrared Processing and Analysis Center/California 
Institute of Technology, funded by the National Aeronautics and Space 
Administration and the National Science Foundation.
We acknowledge useful discussions with Heather Morrison on
the selection and classification of K and M giants.
We acknowledge useful discussions with Jim Gunn 
on the nature and selection of M giants.  We acknowledge several important
observations by the referee, which led to an improved paper.


Funding for the SDSS and SDSS-II has been provided by the Alfred
P. Sloan Foundation, the Participating Institutions, the National
Science Foundation, the U.S. Department of Energy, the National
Aeronautics and Space Administration, the Japanese Monbukagakusho, the
Max Planck Society, and the Higher Education Funding Council for
England. The SDSS Web Site is http://www.sdss.org/.

The SDSS is managed by the Astrophysical Research Consortium for the
Participating Institutions. The Participating Institutions are the
American Museum of Natural History, Astrophysical Institute Potsdam,
University of Basel, Cambridge University, Case Western Reserve University, 
University of Chicago, Drexel University, Fermilab, the
Institute for Advanced Study, the Japan Participation Group, Johns
Hopkins University, the Joint Institute for Nuclear Astrophysics, the
Kavli Institute for Particle Astrophysics and Cosmology, the Korean
Scientist Group, the Chinese Academy of Sciences (LAMOST), Los Alamos
National Laboratory, the Max-Planck-Institute for Astronomy (MPIA),
the Max-Planck-Institute for Astrophysics (MPA), New Mexico State
University, Ohio State University, University of Pittsburgh,
University of Portsmouth, Princeton University, the United States
Naval Observatory, and the University of Washington.

\clearpage

\clearpage

\setcounter{page}{1}
\begin{figure}
\includegraphics[scale=0.9,viewport=-1in -3in 7in 5in]{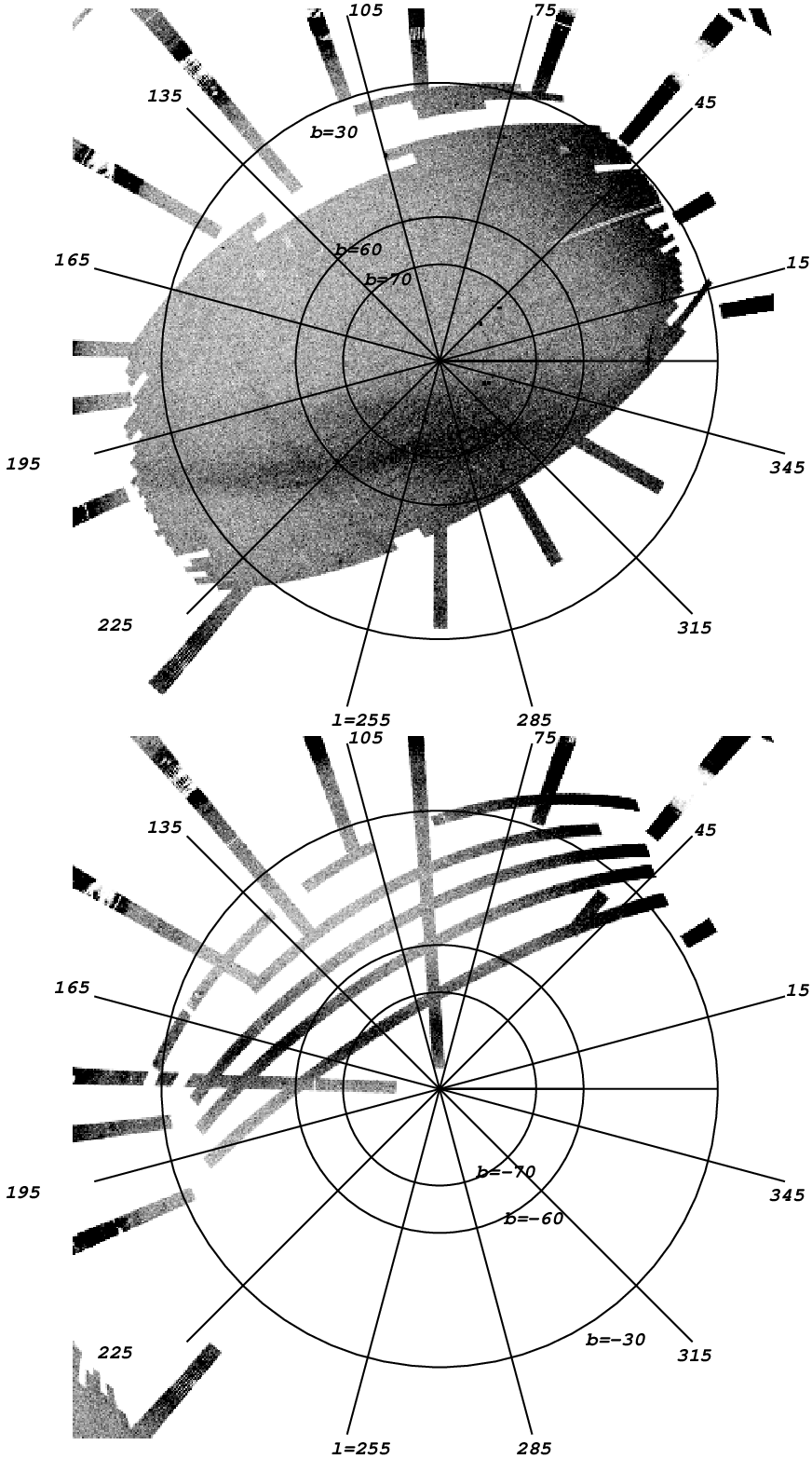}
\caption[Polar density plot of F-turnoff stars]{
\footnotesize
The upper panel shows the density of photometrically selected F-turnoff 
stars with  $21 < g_0 < 22$ in the North Galactic Cap. In addition to the 
prominent Sagittarius stream, note the Orphan stream and Pal~5 are 
visible at $(l,b) = (195^\circ : 255^\circ,50^\circ)$ and $(l,b) = (1^\circ,45^\circ)$,
respectively.  Note that the Sgr leading tidal tail, which dominates the
star counts in this figure, appears to be bifurcated; the main leading
tidal tail is at lower Galactic latitudes and there is a fainter, nearly
parallel tidal tail at higher Galactic latitudes.
The lower panel shows the density of photometrically selected F-turnoff 
stars in the South Galactic Cap.  Note the strong Sagittarius trailing 
tail of F stars running from $(l,b) = (100^\circ,-85^\circ)$ to 
$(170^\circ,-40^\circ)$.  
}
\end{figure}

\begin{figure}
\begin{center}
\includegraphics[scale=0.6]{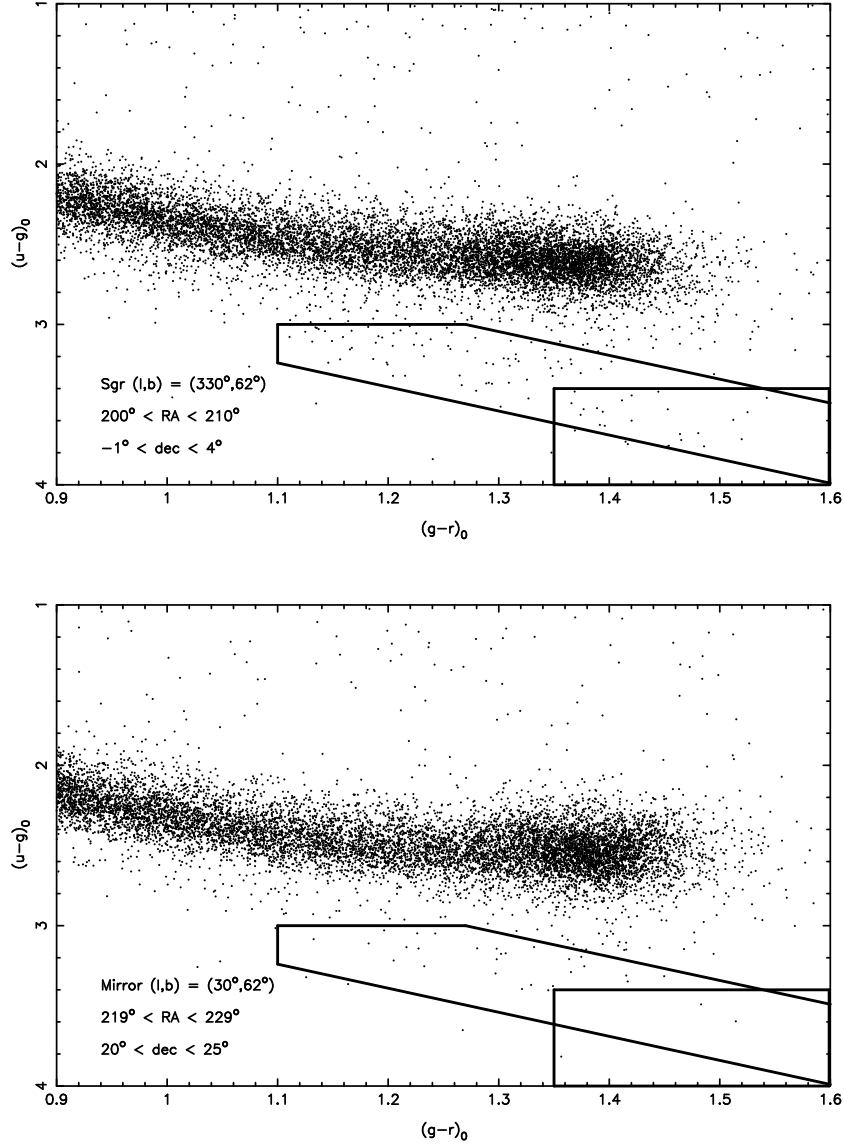}
\end{center}
\caption[Color-color selection of K/M Giants]{
Selection of K/M giants using the fact that they have very red $(u-g)_0$ colors which extend redward of the $(u-g)_0$ M-dwarf stellar locus for $g-r$ redder than 1.1.
Upper panel:  ($(g-r)_0,(u-g)_0$) color-color diagram red stars in 
a 50 square degree box covering the Sagittarius North tidal stream leading
tail.  Note the excess of stars below the locus at red $(u-g)_0$: these
are distant K/M giant candidates.   The outlined selection boxes are 
described in the text.
Lower panel: Same size selection box as above, but in a mirror image 
direction on the sky in Galactic $l$ with respect to the Galactic center.
There are significantly fewer K/M giant candidates, especially in the lower-right
rectangular `very red' selection box.
\label{ugr} }
\end{figure}

\begin{figure} 
\includegraphics[scale=0.7,viewport=-1in 7in 7in 17.0in]{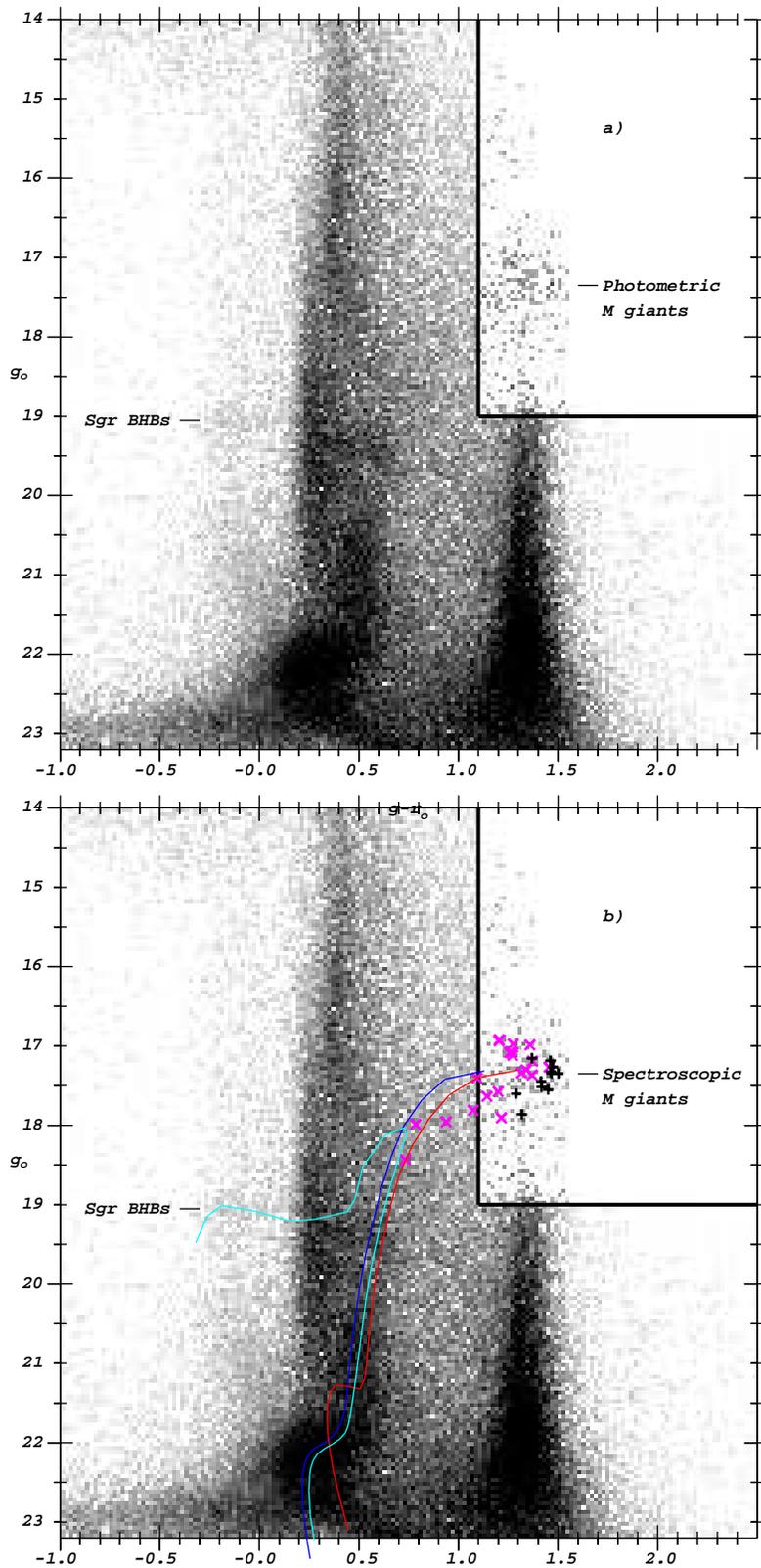}
\caption[Hess diagram Sgr North Stream]{
\footnotesize
The upper panel, a), of this $(g-r)_0,g_0$ color-magnitude Hess diagram is reproduced from \cite{netal02}, Figure 5, updated with DR7 data.  It covers the area ($\alpha,\delta$) = $(200:225^\circ,-4:1^\circ)$.  The inset in the upper right consists only of
candidate K/M-giant stars which pass the strip cut color selection (larger
diagonal box) of Figure 2. 
The relative number of stars in the inset is multiplied by a factor of 10 compared with the main panel to highlight the end of the giant branch.
In the lower panel, b), fiducial sequences from clusters M~3 with [Fe/H] $= -1.5$, (cyan), 
including a BHB
 and M~92 with [Fe/H] $= -2.1$ (blue), are shifted in magnitude
to match the turnoff of Sagittarius stream, and superimposed.
 M~71 with [Fe/H] $= -0.8$ (red) is shifted to match the giant branch.
Spectroscopically confirmed giants are marked with a magenta 
x (bluer K giants, with no TiO bands observed) or with a black + 
(redder M giants, with strong TiO).
\label{hess}
}
\end{figure}

\begin{figure}
\includegraphics[scale=0.65,angle=-90]{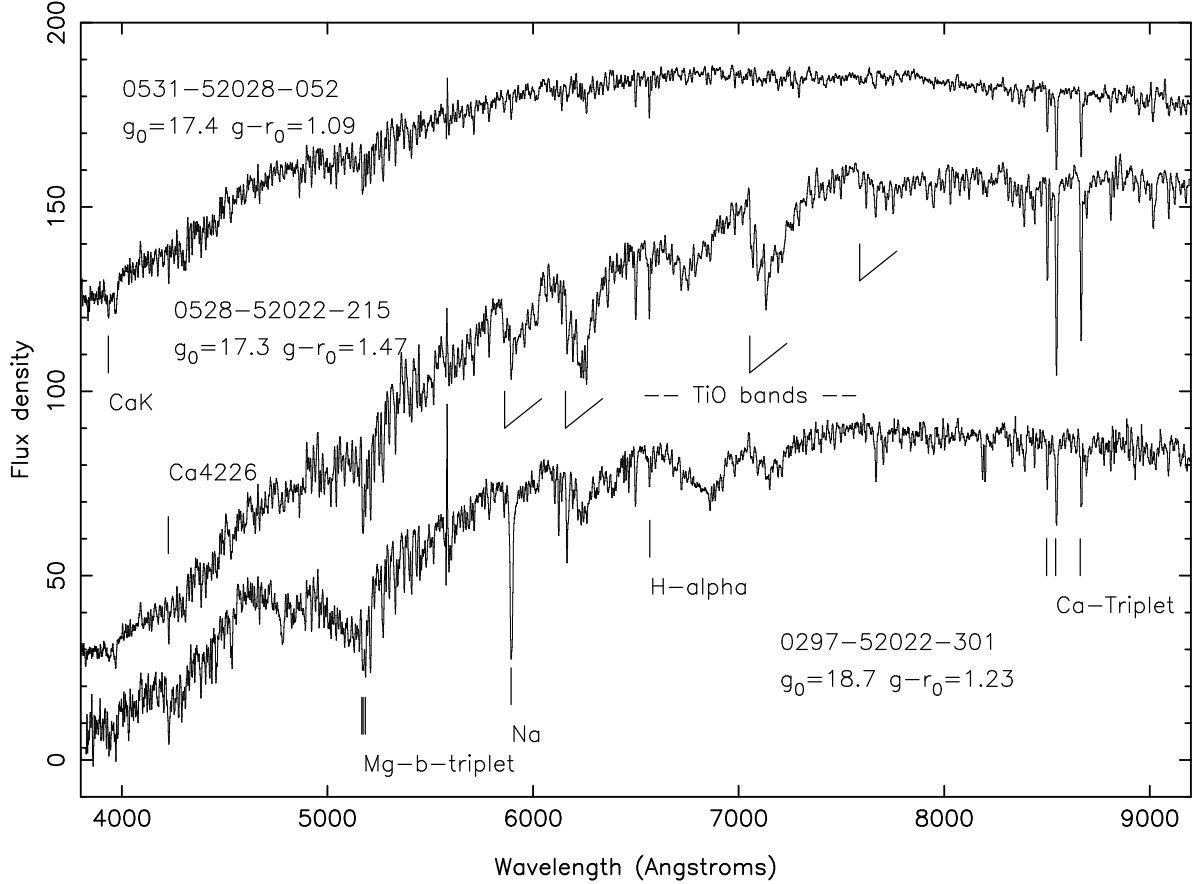}
\caption[Sample K- and M-star spectra]{
Sample K and M star spectra: Three SDSS red-star spectra from the direction of sky indicated in 
Figures 1 and 2.  The upper spectrum (offset by 120 units) is a K giant 
with $(g-r)_0 = 1.09$; note
the weak Mg triplet near 5200\AA\ and the weak Ca 4226\AA . The middle spectrum,
(offset by 20 units) is a very red M giant with $(g-r)_0 = 1.47$; note the
weak Mg triplet, weak Na 5896, yet strong TiO bands.  
The lower spectrum is an M dwarf, with strong Ca 4226, Mg$b$/H, and Na I features.  
The giants are $\sim 44 $ kpc distant
while the dwarf is at about $d=1 $ kpc.\label{samplespecs}}
\end{figure}

\begin{figure}
\includegraphics[scale=0.65,angle=-90]{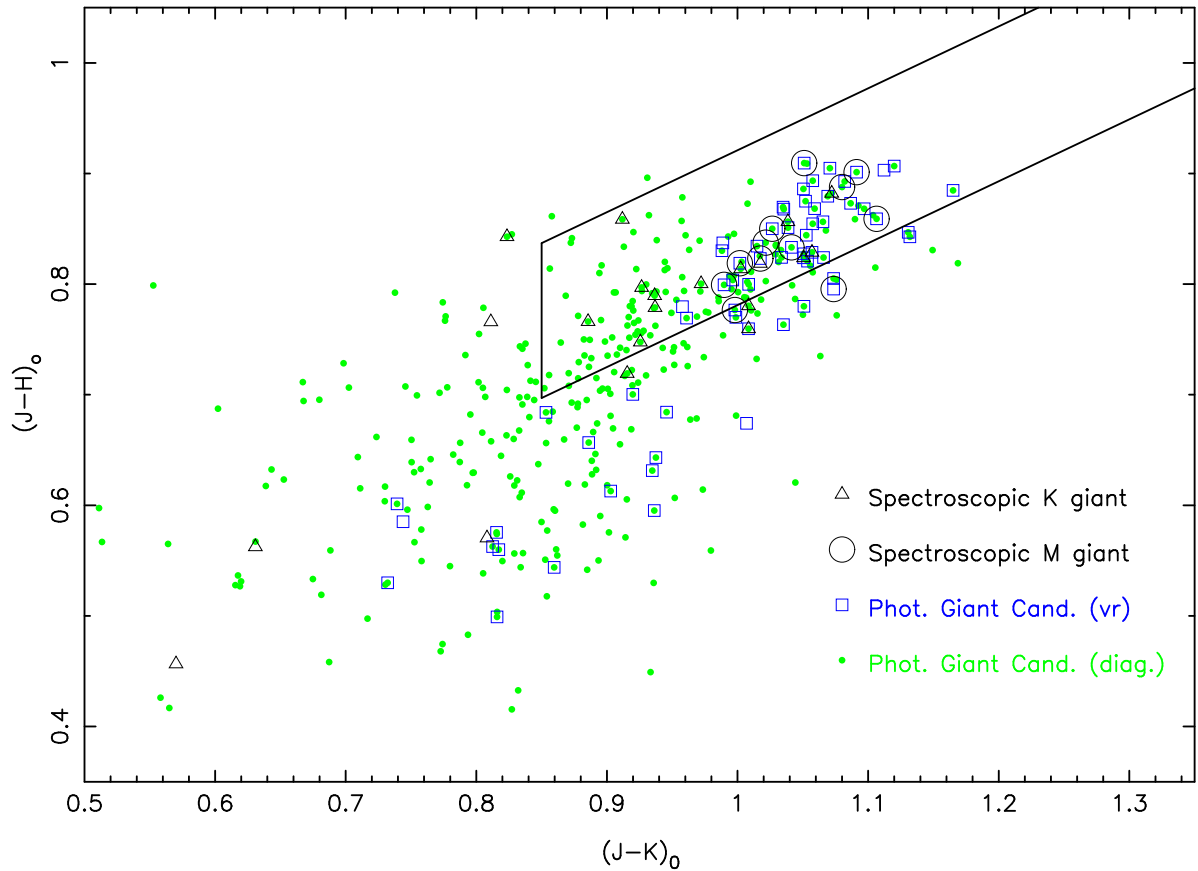}
\caption[2MASS M giant colors]{
A 2MASS IR color-color diagram of Sgr tidal stream K- and M-giant candidates that
were photometrically selected in SDSS $ugr$ filters, as well as spectroscopically confirmed K and M giants.  
The parallelogram box is that of \citet{mswo03}, who originally 
used 2MASS giants to trace the Sgr stream around the Milky Way.  
The large open black circles are SDSS spectroscopically confirmed M giants, the triangles are spectroscopically confirmed K giants (giants with no TiO bands), the filled
(green) circles are Figure 2, upper plot, strip-cut photometric candidates, and the open (blue) squares are
very red $ugr$ candidates from the upper panel in Figure 2. 
Our selection has strong
overlap with the M-giant selection of \citet{mswo03}. 
\label{twomass}} 
\end{figure}

\begin{figure}
\includegraphics[scale=0.9,angle=-90,viewport=0in 2in 6in 11in]{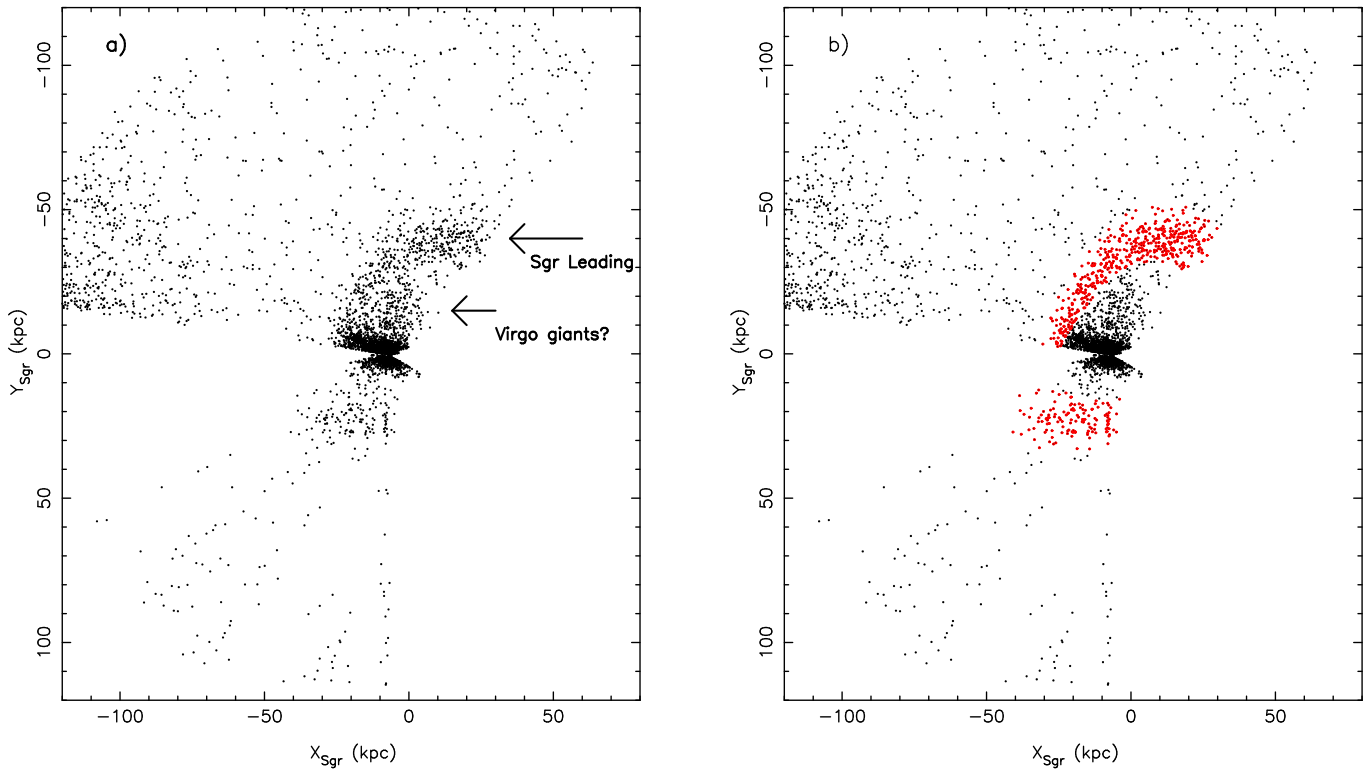}
\caption[Sgr K/M giant selection]{
\footnotesize
The left panel shows the color-selected K/M-giant candidates from all 
SDSS+SEGUE imaging data within 5 kpc of the Sgr dwarf orbital plane.  
The right panel shows the same stars, except the stars selected 
by position in the Sgr leading and trailing tidal streams are colored
red. 
The coordinates are X$_{Sgr}$, Y$_{Sgr}$, and Z$_{Sgr}$ as defined
in \citet{mswo03}.  $Y_{Sgr}$ is close to the $-Z$ axis in Galactic
coordinates.  The axis $+X_{Sgr}$ points approximately in the direction of the Galactic
Center, as viewed from the Sun.  The Galactic plane cuts through the
figure at $Y_{Sgr}=0$.
\label{xym} }
\end{figure}

\begin{figure}
\includegraphics[scale=0.9,viewport=-1in -3in 7in 6in]{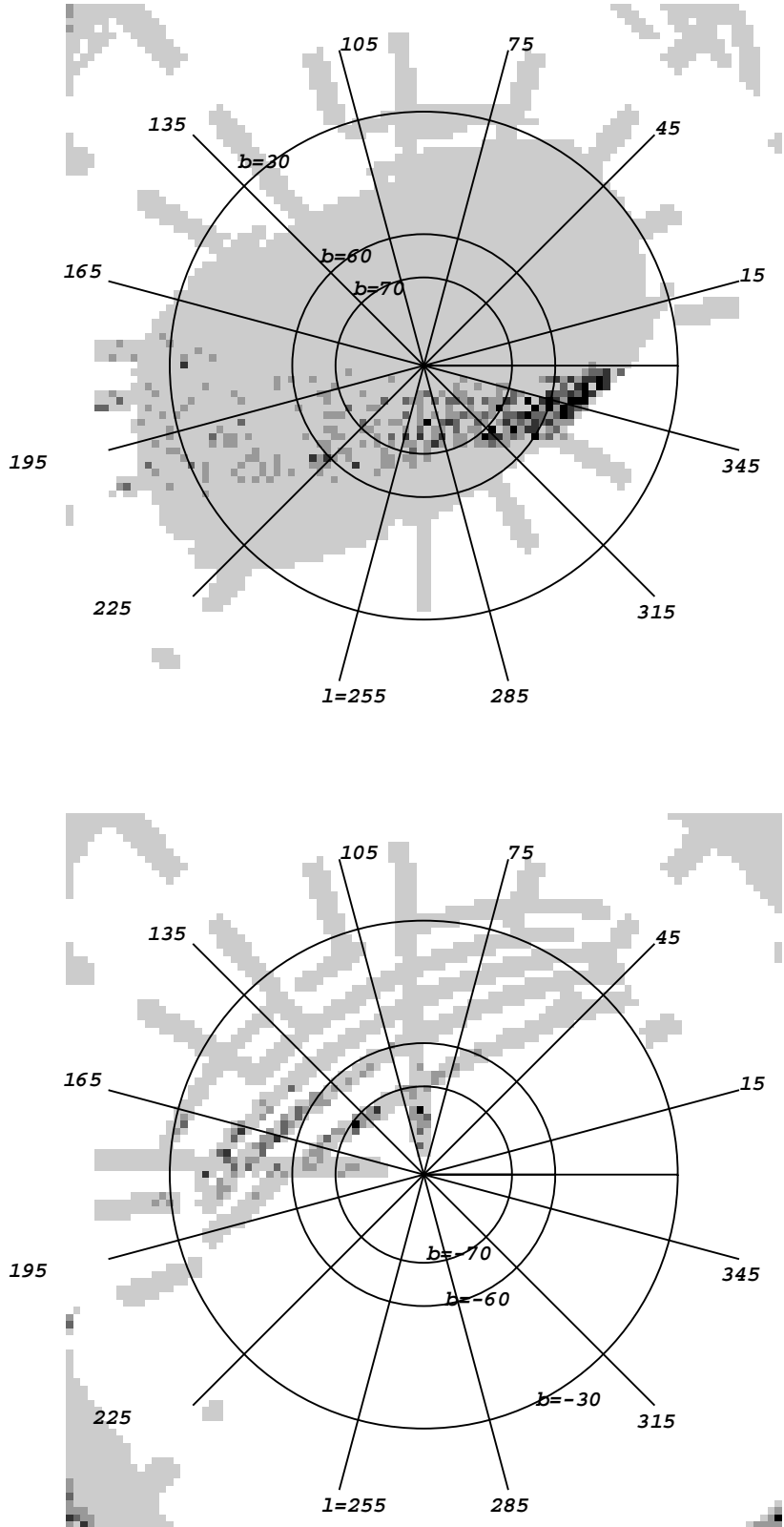}
\caption[Polar density plot of Sgr stream K/M giants] {
\footnotesize
Galactic ($l,b$) polar plots showing the location of Sgr K/M giants colored 
red in Figure 6.  The entire SDSS+SEGUE footprint is indicated in gray.  
The upper panel shows stars in the North Galactic Cap.  Since the
Sgr stream passes closer to the Sun on the left, the angular extent
on the sky is wider there.  The density
of K/M-giant candidates is strongest at the right of the figure,
near $\Lambda_\odot = 300^\circ$, closest to the Sgr 
dwarf itself (not shown).  The lower panel shows stars in the
South Galactic Cap. Note the presence of K/M giants in the Southern
Sgr trailing tidal stream near $l=180^\circ$.
\label{lbpolarm}}
\end{figure}

\begin{figure}
\includegraphics[scale=0.9,angle=-90,viewport=0in 2in 6in 11in]{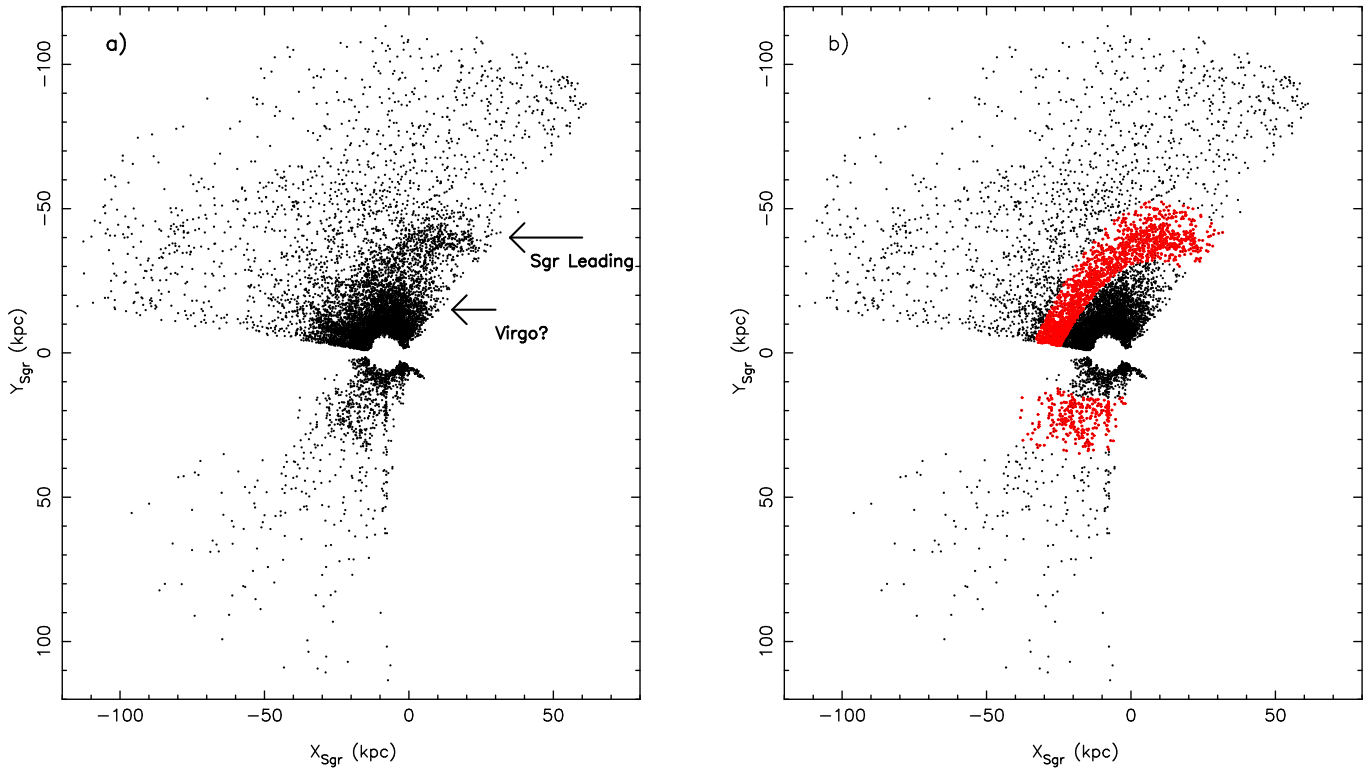}
\caption{Sgr BHB star selection} {
Similar to Figure 6, except photometrically selected BHB
stars are shown instead of K/M giants.  The left panel shows the positions
of BHB stars within 5 kpc of the Sgr stream orbital plane.  
The right panel shows the stars that are candidate Sgr stream stars.  They were selected in exactly the
same positions in the Galaxy as the K/M giants in Figure 6, assuming that
BHB stars are intrinsically 1.7 magnitudes fainter in $g$ than the K/M giants.  Notice that the
background contamination is higher for BHB stars, particularly close to
the Sun.
 }
\end{figure}

\begin{figure}
\begin{center}
\includegraphics[scale=0.9,viewport= -1in -3in 7in 6in]{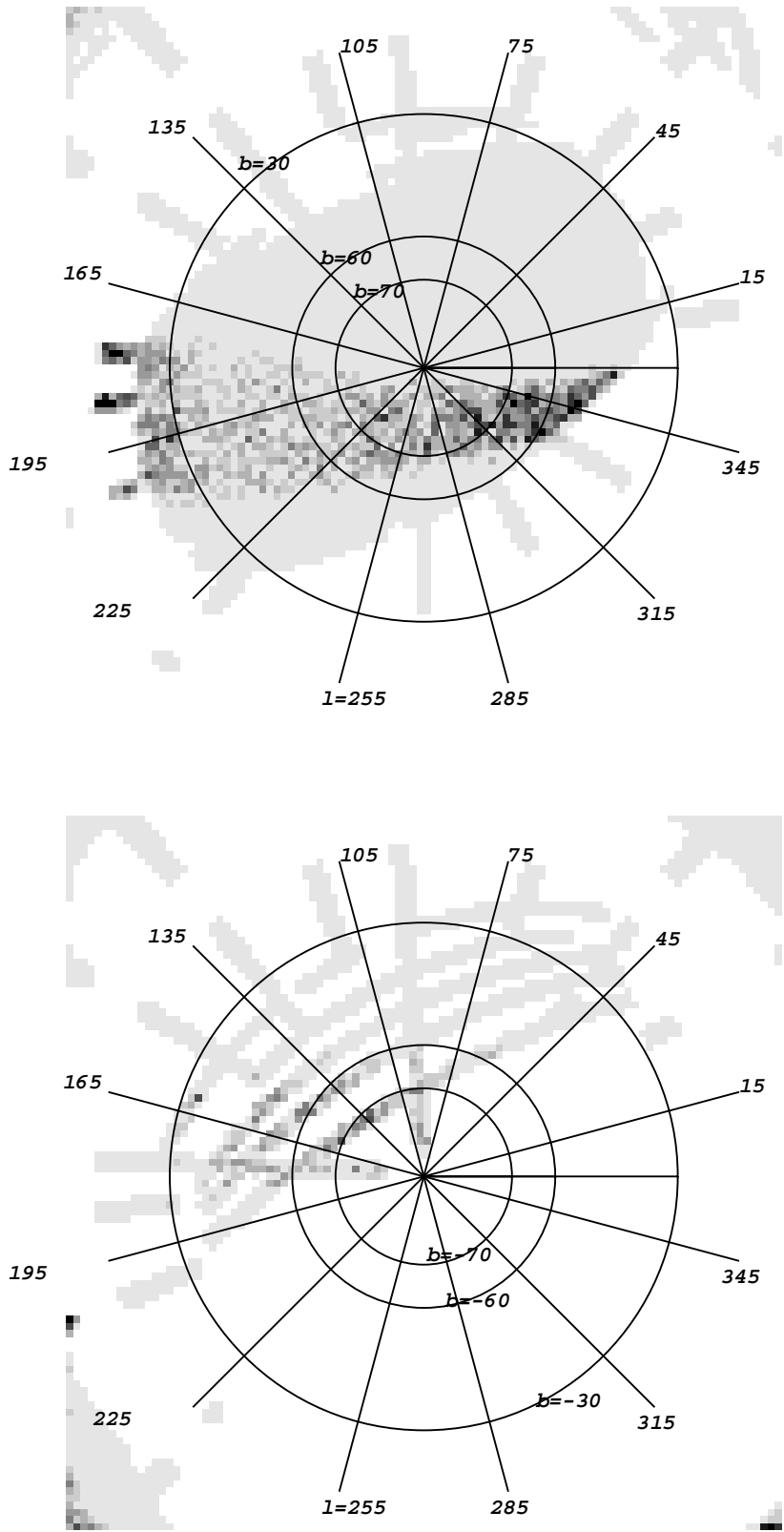}
\end{center}
\caption[Polar density plot of Sgr stream BHB stars] {
\footnotesize
Galactic ($l,b$) polar plots showing the location of Sgr BHBs colored red
in Figure 8.  The entire SDSS+SEGUE footprint
is indicated in gray.  The upper panel shows the  North Galactic Cap and the
lower panel shows the South Galactic pole.  The features are very similar 
to the Figure 7 polar density plot of Sgr stream K/M giants, except that
we see a strong excess of BHB stars near the anticenter in the upper panel.
These stars are not associated with the Sgr dwarf tidal stream.
\label{lbpolara}}
\end{figure}

\begin{figure}
\includegraphics[scale=0.6,viewport=-1in -2in 7in 5in]{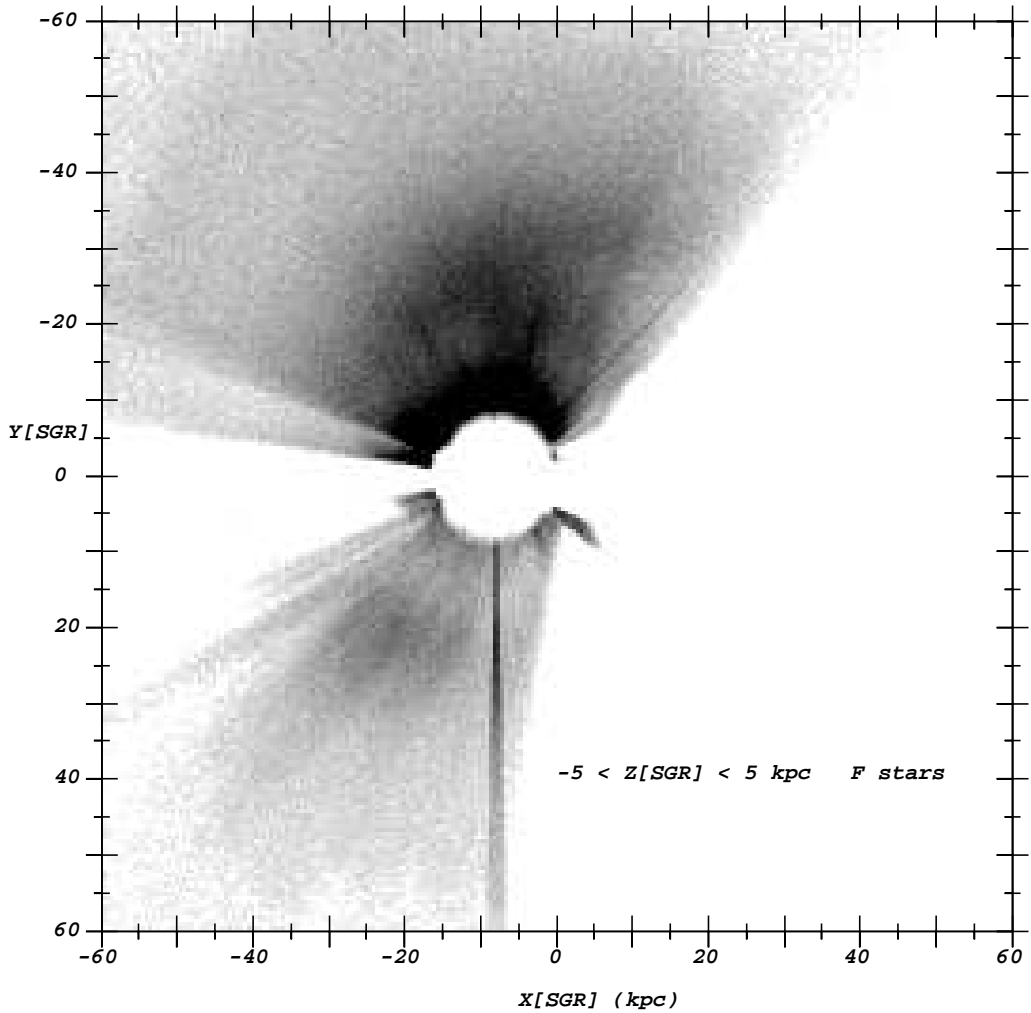}
\caption{Sgr F-turnoff star selection.} {
Similar to Figures 6 and 8, but showing the density of 
photometrically selected F-turnoff stars within 5 kpc of the Sgr dwarf
orbital plane. Since the density of F stars is some
60 times that of BHBs, we use a gray scale density (Hess) diagram rather than plotting
individual stars.  Radial features in this diagram are due to incompleteness,
star clusters, or calibration errors (often due to reddening correction).
The dark radial stripe in the south is due to higher completeness in the
region of a SEGUE imaging stripe at constant Galactic longitude.
We lose completeness of F-turnoff stars at about 40 kpc from the Sun.
in the north, we see the Sgr dwarf leading tidal tail arcing over the
Virgo Overdensity.  We can also see the Sgr trailing tidal tail in the
south.
}
\end{figure}

\begin{figure}
\begin{center}
\includegraphics[scale=0.9,viewport=-1in -2.5in 7in 6in]{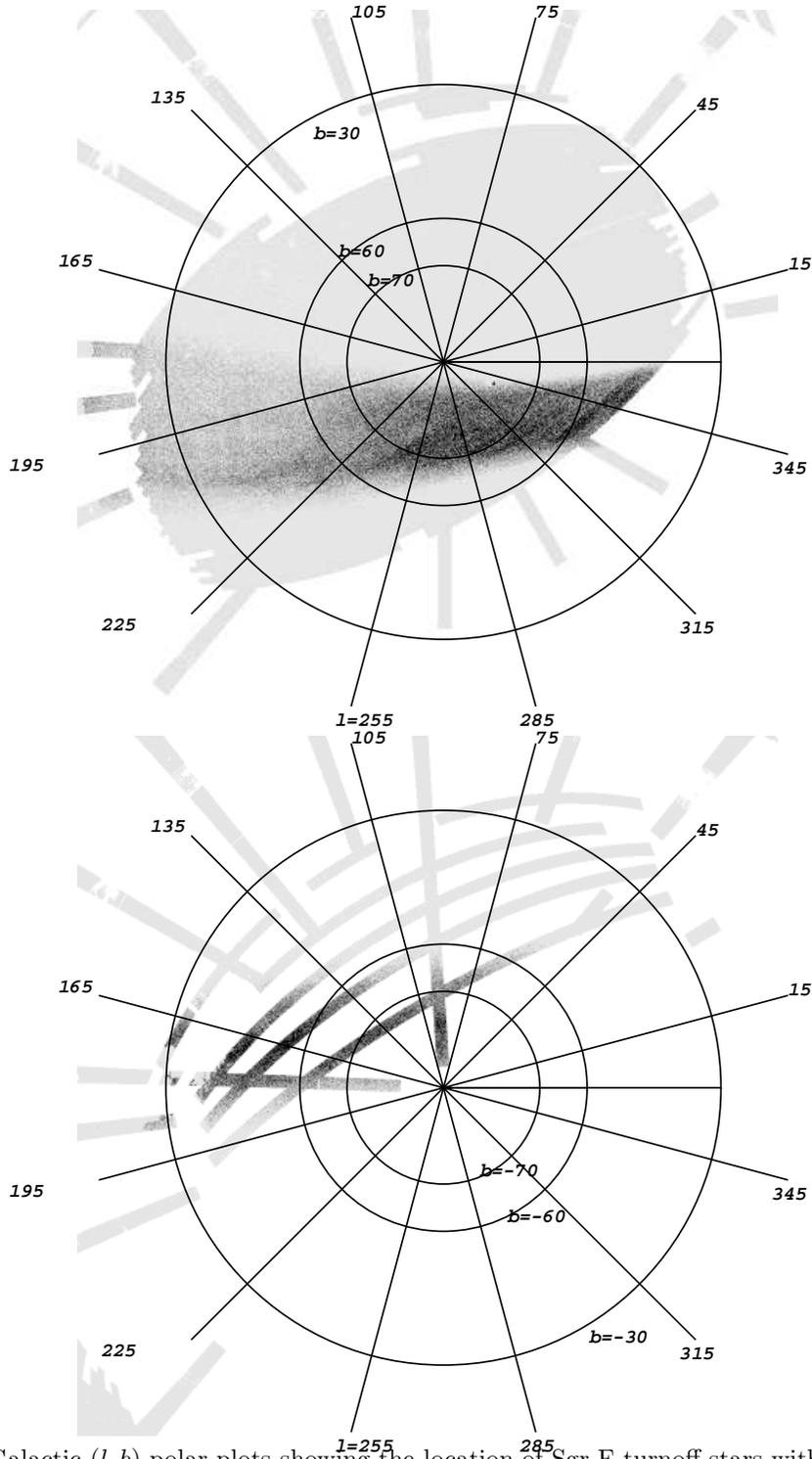}
\end{center}
\caption[Polar density plot of Sgr stream F-turnoff stars] {
\footnotesize
Galactic ($l,b$) polar plots showing the location of Sgr F-turnoff stars 
with $19 < g_0 < 23.5$ and $|Z_{Sgr}| < 5 $ kpc, similar to that for K/M giants and
BHBs in in Figures 6 and 8, assuming all of the F-turnoff 
stars are 5.2 magnitudes fainter than the K/M-giant stars.  
Since there is actually a very broad range of
F-turnoff star absolute magnitudes, some stars may be missed in this
selection.  
The entire SDSS+SEGUE footprint
is indicated in gray.  The upper panel shows stars in the  North Galactic 
Cap, and the lower panel shows stars in the South Galactic Cap.
}
\end{figure}

\begin{figure}
\begin{center}
\includegraphics[scale=0.7,viewport=1in 0in 6in 9in]{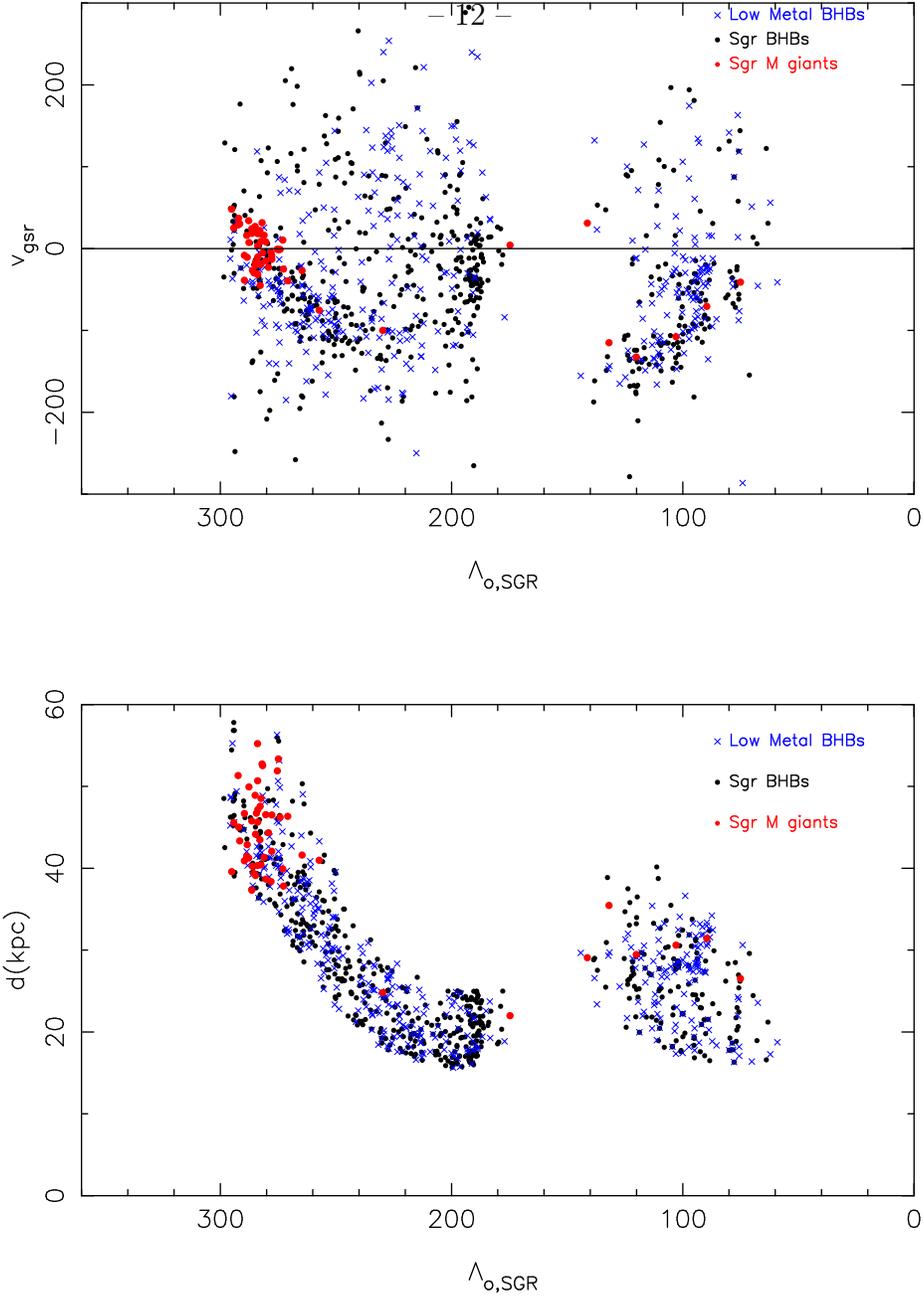}
\end{center}
\caption[Velocities of tidally stripped Sgr stars] {
\footnotesize
The top panel shows line-of-sight, Galactic standard of rest velocities, 
$v_{gsr}$, for all HB stars with SDSS/SEGUE spectra (selected as stars with 
$-0.3 < (g-r)_0 < 0.35$ and $1 < \rm log~g < 3.75$) that are within
5 kpc of the Sgr dwarf orbital plane, and spatially coincident with 
the Sgr stream.  We also show those photometrically selected K/M giant stars
for which with SDSS+SEGUE spectra were taken (not many, and over a small
area of sky).  This plot mirrors that of
Figure 12 in \citet{ljm05}.   The angle $\Lambda_{\odot,SGR}$ increases along
the Sgr tidal stream, as defined in \citet{mswo03}.  The HB stars are 
divided into low metallicity ($\rm [Fe/H]_{\rm WBG} < -1.9$, indicated with blue crosses),  
and high metallicity ($\rm [Fe/H]_{\rm WBG} > -1.9$, filled black dots), samples.  
We find K/M giants (red dots) and both low- and high-metallicity BHB stars in the Sgr
dwarf tidal stream.  BHBs in the Virgo overdensity are apparent at 
$\Lambda_\odot=230^\circ, v_{gsr}=130$ km s$^{-1}$.  A new stellar stream
is apparent in low-metallicity BHB stars near $\Lambda_\odot=100^\circ$,
$v_{gsr}=-50$ km s$^{-1}$.  The lower plot shows the estimated distances
for each of the stars in the upper plot as a function of $\Lambda_\odot$.
The newly identified low-metallicity stream has a very narrow 
distance distribution that rises toward the right in the lower panel.
}
\end{figure}

\begin{figure}
\includegraphics[scale=0.7]{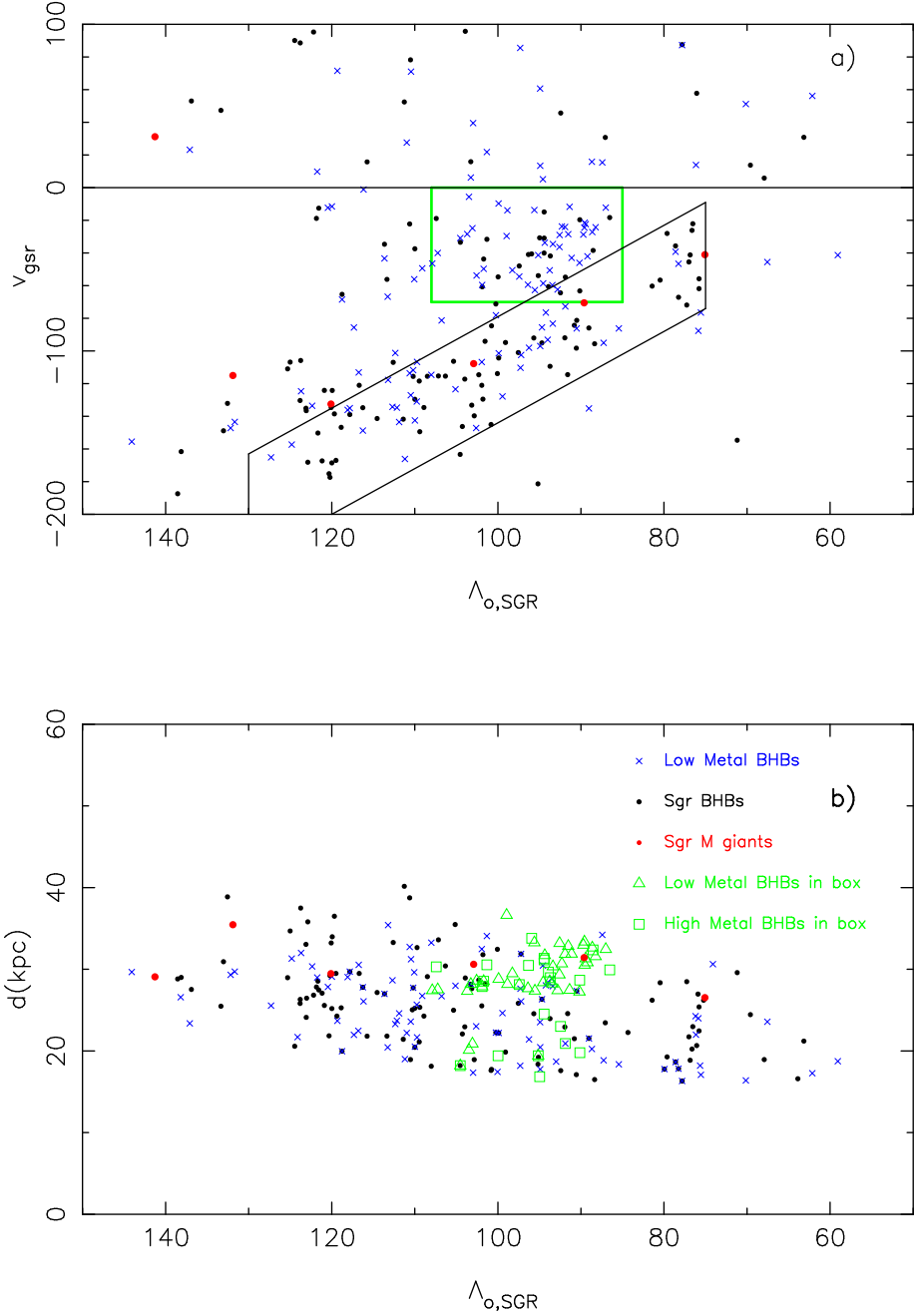}
\caption[Figure 13]{
Same as Figure 12, except focusing on the South where a possible
new stream is visible, distinct from the Sgr trailing tidal tail. 
The rectangular box in the upper panel, a), outlines the area of sky where 
the new stream is prominent in low-metallicity BHB stars.  
The diagonal box shows where the Sgr south trailing tidal tail 
is most prominent, containing stars of mixed metallicity.  Note the ratio
of low- to high-metallicity BHB stars is very different between the
two outlined regions.  
In the lower panel, b), distances to all BHB stars from the rectangular box in the
upper panel are plotted in green triangles (low metallicity) and squares (high metallicity).  Note that these BHBs have a different distance 
distribution with $\Lambda_\odot$ than the Sgr tidal tail stars.
}
\end{figure}

\clearpage

\begin{figure}
\begin{center}
\includegraphics[scale=0.9,viewport=-1in -2.5in 7in 4in]{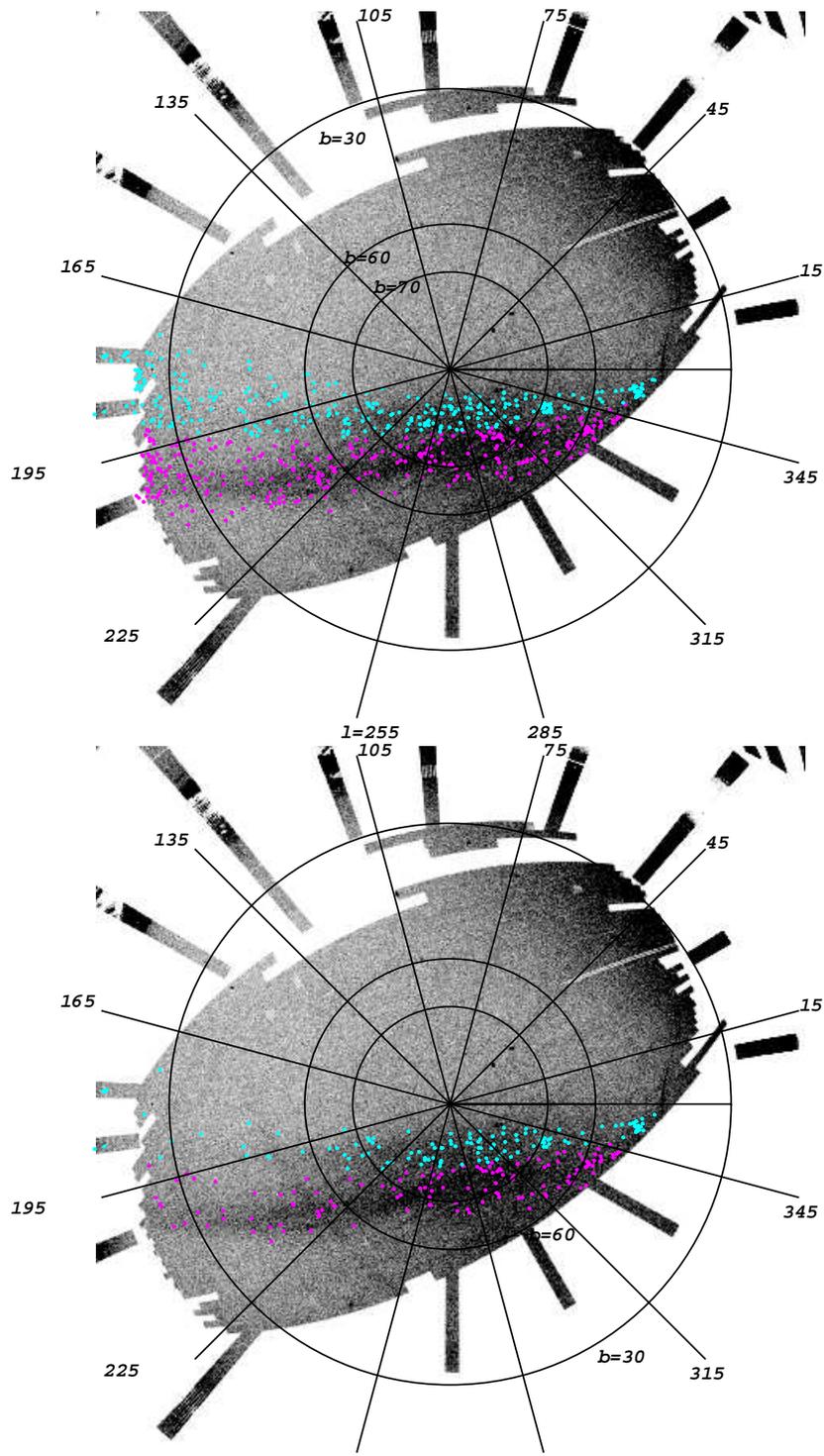} 
\end{center}
\caption[HB spectra in the Sgr leading tidal tail] {
\footnotesize
The upper panel shows the available HB spectra that are spatially coincident with 
the Sgr leading tidal tail; spectra that are candidates for the upper branch 
are shown in cyan and candidates for the lower branch are shown in magenta.
The lower panel shows the subset of the HB spectra that have the correct
velocities to be Sgr dwarf tidal debris.  The sparse sampling of the lower,
main leading tidal tail is a selection effect.  The $(l,b)$ locations of
the spectra with the correct velocities for Sgr stream stars (shown in the lower panel) 
suggest that the two stream branches grow further 
apart (in angle) as they approach the Galactic anticenter.
}
\end{figure}

\begin{figure}
\begin{center}
\includegraphics[scale=0.7]{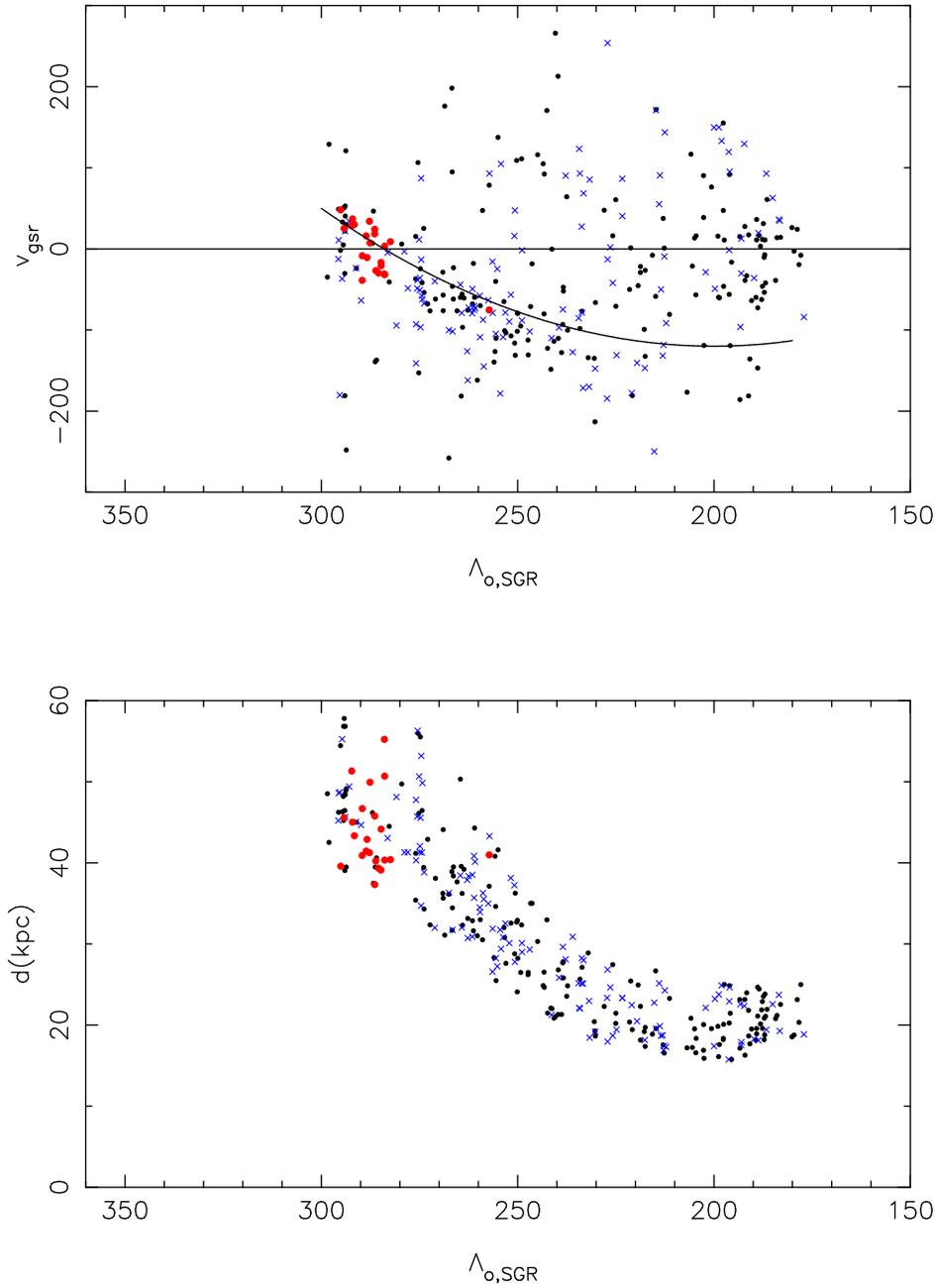}
\end{center}
\caption{Velocities of Sgr leading tidal tail, upper branch.} {
Subsets of stars in Figure 12 that are in the upper branch
of the Sgr leading tidal tail, with $Z_{Sgr}>0.04 X_{Sgr}+1.0$, are shown.  Two lines
that approximately trace the velocities of the Sgr dwarf tidal debris
are shown for reference.
}
\end{figure}

\begin{figure}
\begin{center}
\includegraphics[scale=0.7]{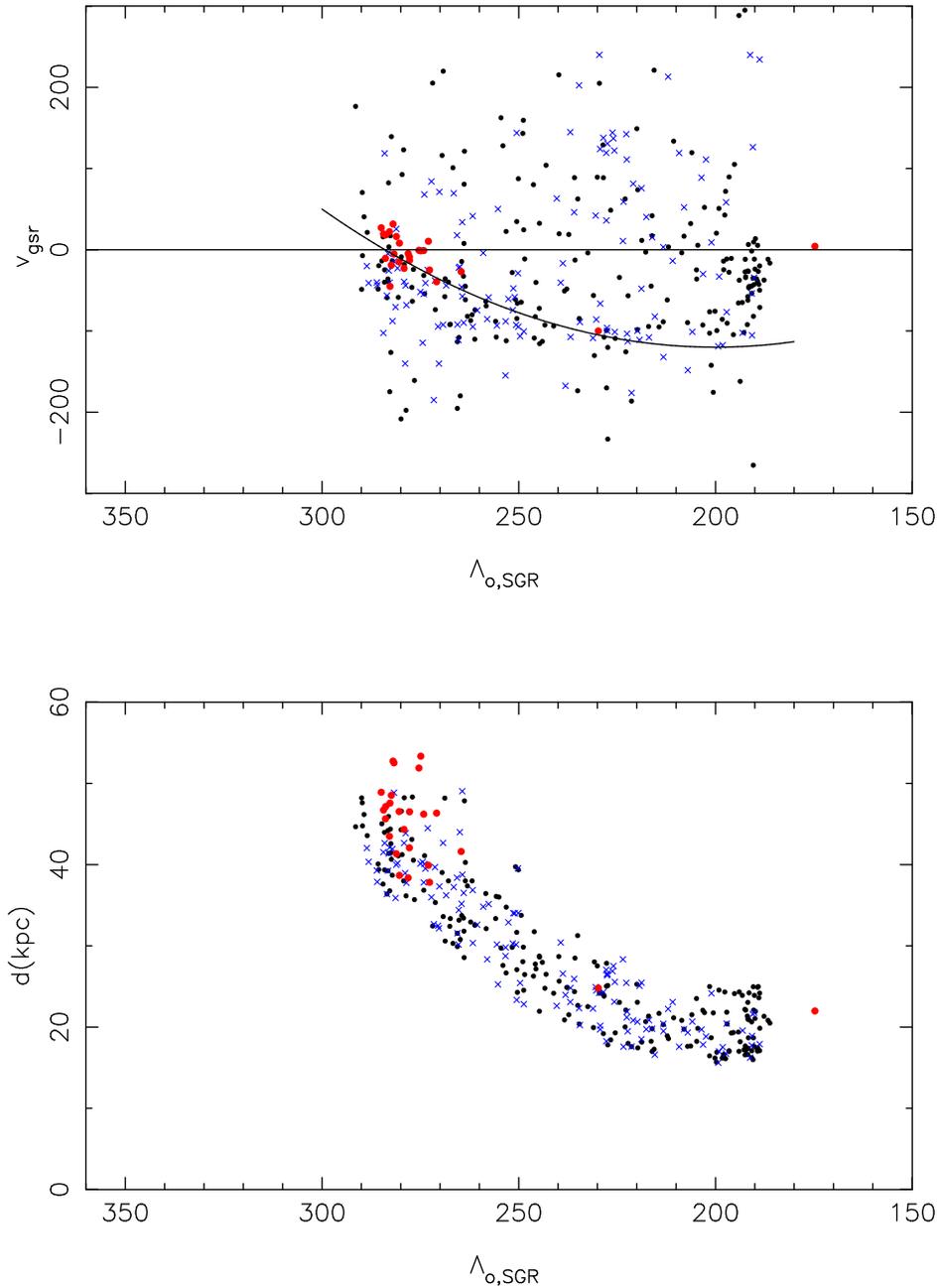}
\end{center}
\figcaption{Velocities of Sgr leading tidal tail, lower branch.} {
Subsets of stars in Figure 12 that are in the lower, main branch
of the Sgr leading tidal tail, with $Z_{Sgr}<0.04 X_{Sgr}+1.0$, are shown.  Two lines
that approximately trace the velocities of the Sgr dwarf tidal debris
are shown for reference.  Note that the velocity distribution appears to
be identical to the velocity distribution for the upper branch (Figure 15).
The K/M giants at slightly different $\Lambda_\odot$ due to a selection effect
(we have K/M-giant spectra over a very small area of the sky).
}
\end{figure}

\begin{figure}
\begin{center}
\includegraphics[scale=0.7]{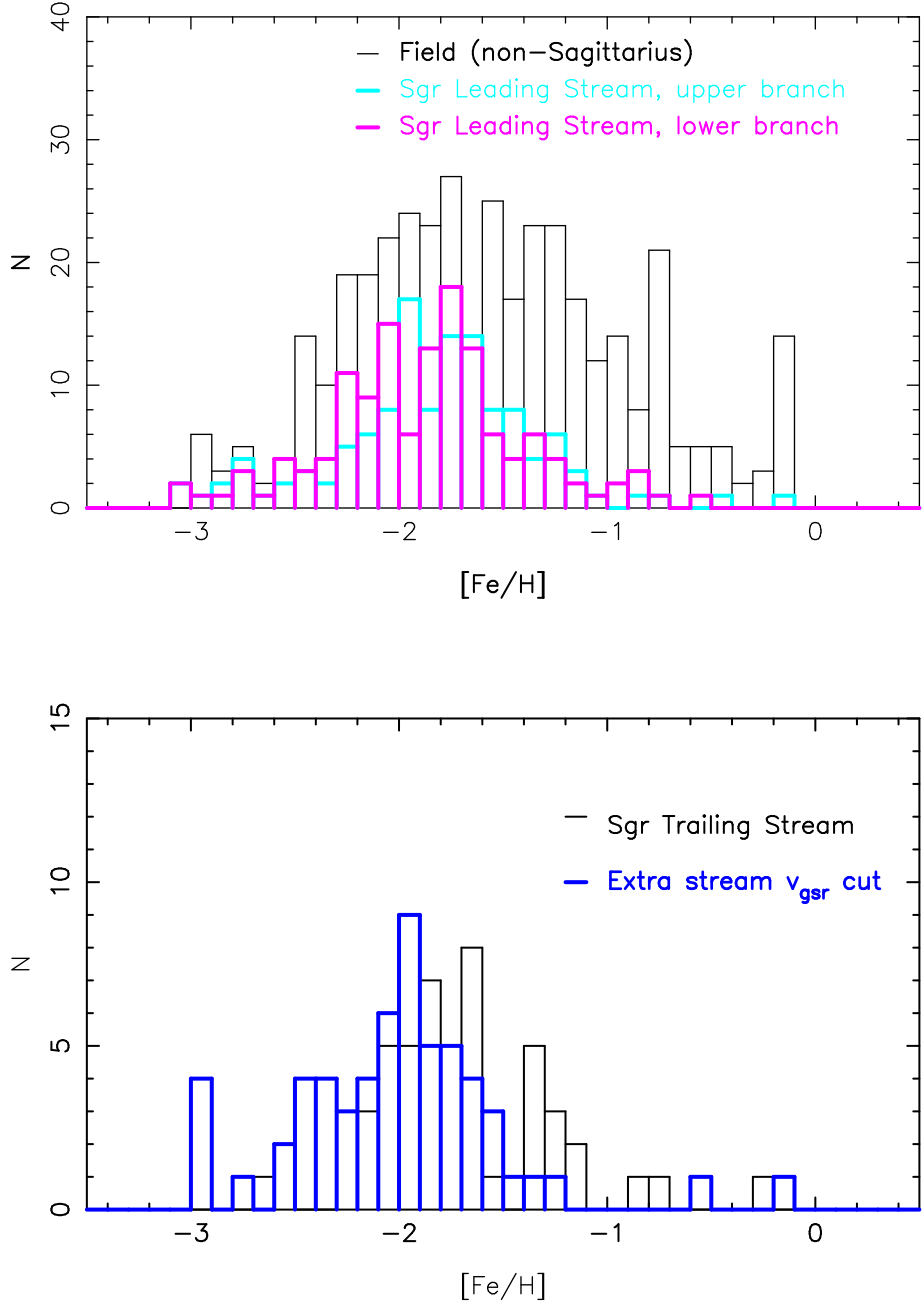}
\end{center}
\caption[Metallicities of HB stars in the Sgr tidal tails.] {
The upper panel shows the metallicity distribution of spectroscopically
selected stars in the Sgr leading tidal tail upper branch (cyan) and 
lower branch (magenta).  In black we show the distribution of metallicities
for all spectroscopically selected HB stars that are spatially coincident
with the Sgr stream, but do not have the correct velocities to be Sgr
tidal debris.  There is no significant distinction between the upper and lower
branch metallicity distributions.  The lower panel shows the metallicity distribution
of the HB stars in the trailing tidal tail (black) along with the metallicity
distribution of the HB stars in the newly discovered tidal stream (selected
with $ -71 < v_{\rm gsr} < 0$ km s$^{-1}$
and $85^\circ<\Lambda_\odot < 130^\circ$).  The metallicities in the
new stream are lower than the average Sgr HB star.  The spikes for HB metallicities
at [Fe/H] = $-3.0$ are spurious.
}
\end{figure}

\begin{figure}
\begin{center}
\includegraphics[scale=0.65,viewport=1in 0.5in 7in 9in]{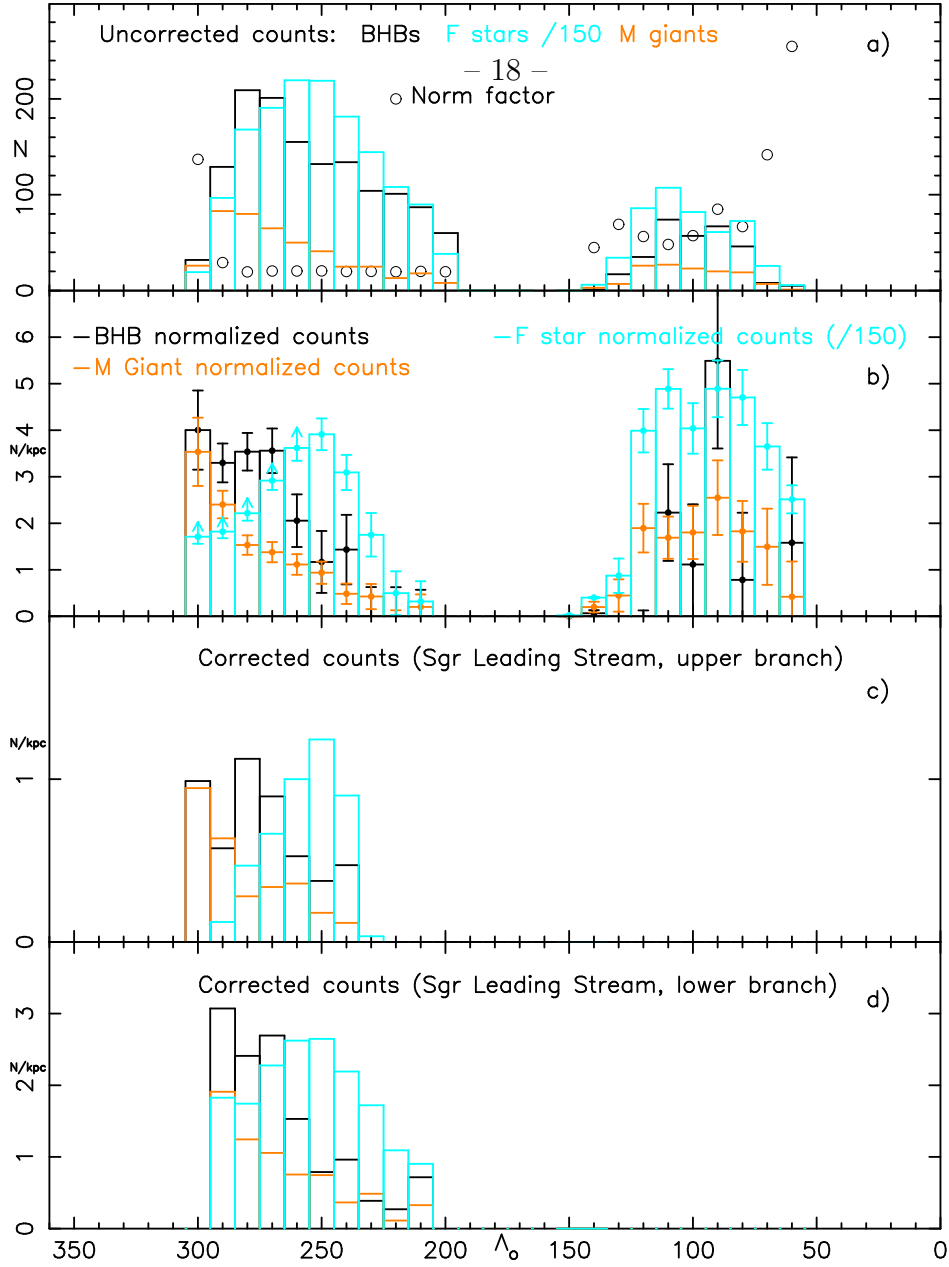}
\end{center}
\caption[Density of K/M giants, HB stars, and F-turnoff stars as a function 
angle along the Sgr stream] {
\footnotesize
The top panel, a), shows the raw number counts
of stars in each photometrically selected category, in ten degree angular
bins along the Sgr dwarf tidal stream in $\Lambda_\odot$ (F-star counts are divided by
150 in all panels).  
The black open circles
indicate the normalization factor (multiplied by 20 so that it is visible
on the plot) that the bin will be multiplied by to account for the
incompleteness of the data in that bin.  
The second panel, b),  shows the star
counts after they have been multiplied by the normalization factor,
background subtracted, and divided by the distance to the Sgr tidal stream
at that angular position in the sky.  
Note that since the F-turnoff stars
in the Sgr stream are too faint to be observed in SDSS data in the first
five bins, we have only lower limits for the F-turnoff star number counts
on the left side of the plot.  Notice that the ratio of K/M-giant stars to
HB stars to F-turnoff stars is constant within the errors of the plot.
The density of the leading tidal tail decreases from the left side of
the plot toward decreasing $\Lambda_\odot$.  The density along the observed
portion of the Sgr trailing tidal tail is roughly constant.  The lower
two panels show the data in the leading tidal tail, split by upper, c), 
and lower, d), branches.  The normalization for each bin was recalculated, and the subtracted
background was scaled by the fraction of the stream width that is in 
each branch in each bin.  There is no apparent difference in the ratio of
K/M giants to BHB stars to F-turnoff stars.  The upper leading tidal tail
has less than half the density, and appears to end abruptly at $\Lambda_\odot \sim 235^\circ$,
however we note that bins where the background subtraction results in negative stream counts
have been supressed.
\label{rho}}
\end{figure}

\begin{figure}
\begin{center}
\includegraphics[scale=0.6]{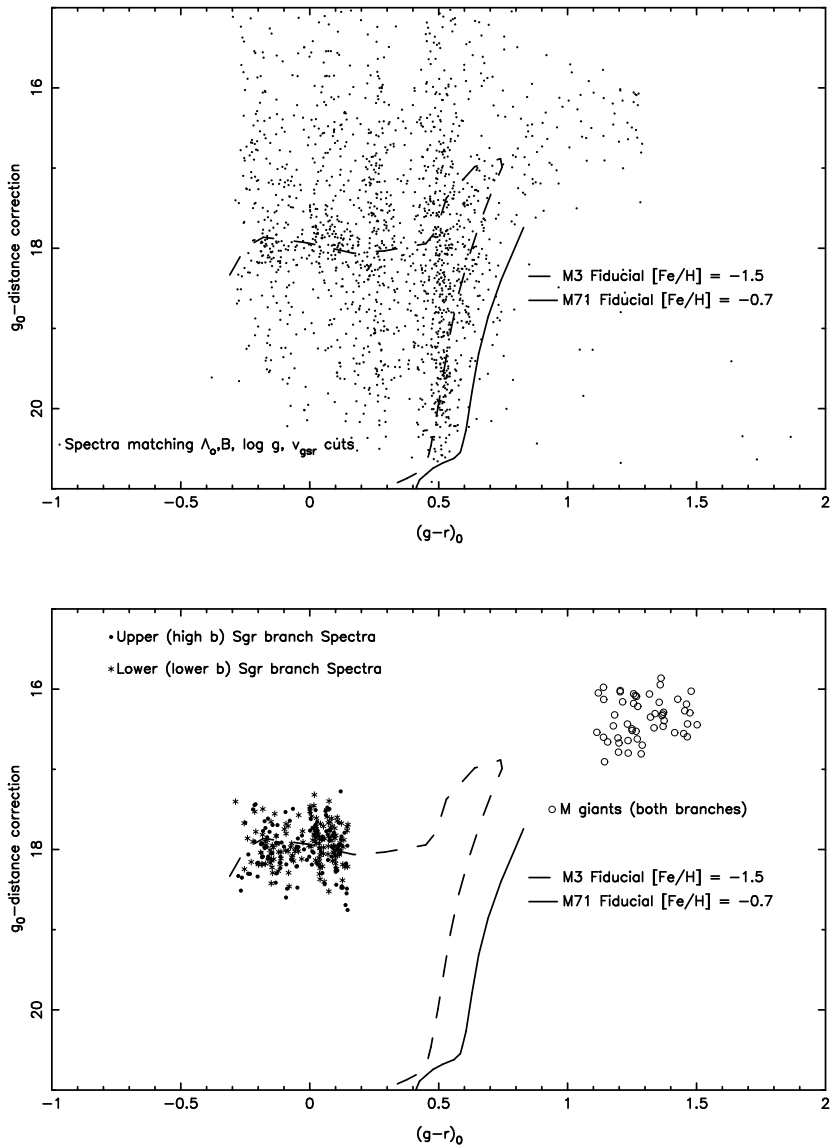}
\end{center}
\caption[Color-magnitude diagram of spectra in Sgr tidal stream.] {
\footnotesize
We selected all stellar spectra in SDSS DR7 within the angular limits of the
Sgr leading tidal tail that have low surface gravity and velocities consistent
with Sgr tidal debris.  We adjusted the apparent magnitude of each star,
based on its $\Lambda_\odot$ coordinate, so that if it is at the distance
of the Sgr stream in that direction it would be shifted to 30 kpc.  This
should give us a Sgr blue horizontal branch at an adjusted $g_0$ of 18.1
and an K/M-giant branch at an adjusted $g_0$ of 16.4.  In the top panel,
we show the adjusted color-magnitude diagram for these stars, which does show the
horizontal branch and K/M-giant branch of Sgr.  The large number of counts
at $(g-r)_0=0.5$ is due to the color selection for SEGUE spectra that favors
this color range.  The fiducial sequences for M~3 and M~71 \citep{anetal08}, 
shifted to 30 kpc,
are shown for reference.  The lower panel shows the Sgr HB and K/M giant stars
from Figure 12 that are at the correct velocities for Sgr tidal debris.
The BHB selection is very good.  Since $\rm log~ g$ is not measured reliably for M giants
with $(g-r)_0 > 1.3$ in SDSS, they cannot be spectroscopically selected and are therefore
absent in the top panel.  Our lower panel K/M giant candidates are spectra that were
selected by their photometric indices.
}
\end{figure}










\clearpage
\begin{thebibliography}{}
\bibitem[Abazajian et al.(2009)]{aetal08} Abazajian K., et al. 2009, ApJS, 182, 543.
\bibitem[Abazajian et al.(2003)]{aetal03} Abazajian et al. 2003, AJ, 126,2081
\bibitem[Allende Prieto et al.(2008)]{apetal08} Allende Prieto, C. et al. 2008 AJ, 136, 2070.
\bibitem[Allgood et al.(2006)]{allgood06} Allgood, B., Flores, R., Primack, J. Kravtsov, A. Wechsler, R., Faltenbacher, A. \& Bullock, J. 2006 MNRAS 367, 1781.
\bibitem[An et al.(2008)]{anetal08} An, D., Johnson, J. et al.  2008 ApJS, 179, 326.
\bibitem[Bell et al.(2008)]{betal07} Bell, E. F. et al. 2008, ApJ, 680, 295.
\bibitem[Bellazini et al.(2006)]{betal06} Bellazzini, M., Newberg, H. J., Correnti, M., Ferraro, F. R., \& Monaco, L. 2006, \aa, 457, L21
\bibitem[Belokurov et al.(2006a)]{2006ApJ...637L..29B} Belokurov, V., Evans, N.~W., Irwin, M.~J., Hewett, P.~C., \& Wilkinson, M.~I.\ 2006, \apjl, 637, L29 
\bibitem[Belokurov et al.(2006b)]{betal06a} Belokurov, V. et al. 2006a, ApJ 642, L137.
\bibitem[Belokurov et al.(2006c)]{betal06b} Belokurov, V. et al. 2006b, ApJ 647, L111.
\bibitem[Belokurov et al.(2007)]{betal07a} Belokurov, V. et al. 2007a, ApJ 658,337 
\bibitem[Belokurov et al.(2007)]{2007ApJ...657L..89B} Belokurov, V., et 
al.\ 2007, \apjl, 657, L89 
\bibitem[Chou et al.(2007)]{chouetal07} Chou, M.-Y. et al. 2007, \apj, 670, 346
\bibitem[Clem, Vanden Berg, \& Stetson(2008)]{clemetal08} Clem, J. L., Vanden Berg, D. A., Stetson, P. B. 2008 AJ 135, 682
\bibitem[Dehnen \& Binney(1998)]{db98} Dehnen, W. \& Binney, J. J. 1998, \mnras, 298, 387
\bibitem[Duffau et al.(2006)]{detal06} Duffau, S., Zinn, R., Vivas, A. K., Carraro, G., Mendez, R. A., Winnick, R., \& Gallart, C. 2006, \apj, 636, L97
\bibitem[Fellhauer et al.(2006)]{fetal06} Fellhauer, M., et al. 2006, \apj, 651, 167
\bibitem[Fukugita et al.(1996)]{figdss96} Fukugita, M., Ichikawa,T., Gunn, J. E., Doi, M., Shimasaku, K., Schneider, D. P. 1996, \aj, 111, 1758
\bibitem[Grillmair(2008)]{2008arXiv0811.3965G} Grillmair, C.~J.\ 2008, arXiv:0811.3965
\bibitem[Grillmair(2006a)]{2006ApJ...645L..37G} Grillmair, C.~J.\ 2006a, \apjl, 645, L37 
\bibitem[Grillmair(2006b)]{2006ApJ...651L..29G} Grillmair, C.~J.\ 2006b, \apjl, 651, L29 
\bibitem[Grillmair \& Dionatos(2006)]{2006ApJ...643L..17G} Grillmair, C.~J., \& Dionatos, O.\ 2006, \apjl, 643, L17 
\bibitem[Grillmair \& Johnson(2006)]{2006ApJ...639L..17G} Grillmair, C.~J., \& Johnson, R.\ 2006, \apjl, 639, L17 
\bibitem[Gunn et al.(1998)]{getal98} Gunn, J. E. et al. 1998, \aj, 116, 3040
\bibitem[Gunn et al.(2006)]{getal06} Gunn, J. E. et al. 2006, \aj, 131, 2332
\bibitem[Harrigan et al.(2009)]{harrigan} Harrigan, M. J., Newberg H. J., Yanny, B., Beers, T.C., Lee, Y.S., \& Re Fiorentin, P., in preparation
\bibitem[Helmi et al.(2003)]{hetal03} Helmi, A. et al. 2005, \apj, 586, 195
\bibitem[Helmi(2004)]{h04} Helmi, A. 2004, \apj, 610, L97
\bibitem[Hogg et al.(2001)]{hfsg01} Hogg, D. W., Finkbeiner, D. P., Schlegel, D. J., \& Gunn, J. E. 2001, \aj, 122, 2129
\bibitem[Ibata, Gilmore, and Irwin(1994)]{ietal94} Ibata, R. A., Gilmore, G., and Irwin, M. J. 1994, \nat, 370, 194
\bibitem[Ibata et al.(2001)]{ietal01} Ibata, R., Lewis, G. F., Irwin, M., Totten, E., and Quinn, T. 2001, \apj, 551, 294
\bibitem[Ibata et al.(2003)]{ietal03} Ibata, R. A., Irwin, M. J., Lewis, G. F., Ferguson, A. M. N., \& Tanvir, N 2003, MNRAS 340, L21
\bibitem[Ivezi\'{c} et al.(2000)]{ietal00} Ivezi\'{c}, Z., et al. 2000, \aj, 120, 963
\bibitem[Ivezi\'{c} et al.(2004)]{ietal04} Ivezi\'{c}, Z., et al. 2004, Astronomische Nachrichten, 325, 583
\bibitem[Johnston, Law \& Majewski(2005)]{jlm05} Johnston, K. V., Law, D. R., \& Majewski, S. R. 2005, \apj, 619, 800
\bibitem[Juri\'{c} et al.(2008)]{jetal06} Juric, M., et al. 2008, ApJ 673, 864.
\bibitem[Keller, Da Costa, \& Prior(2009)]{kdp09} Keller, S. C., Da Costa, G. S., \& Prior, S. L. 2009 MNRAS, 394, 1045.
\bibitem[Layden et al.(1996)]{letal96} Layden, A.~C., Hanson, R.~B., Hawley, S.~L., Klemola, A.~R., \& Hanley, C.~J.\ 1996, \aj, 112, 2110 
\bibitem[Law, Johnston, \& Majewski(2005)]{ljm05} Law, D. R., Johnston, K. V. \& Majewski, S. R. 2005, \apj, 619, 807
\bibitem[Lee et al.(2008a)]{letal08a} Lee, Y. S. et al. 2008 AJ 136, 2022.
\bibitem[Lee et al.(2008b)]{letal08b} Lee, Y. S. et al. 2008 AJ 136, 2050.
\bibitem[Lenz et al.(1998)]{lnrrs98} Lenz, D. D., Newberg, H. J., Rosner, R., Richards, G. T., \& Stoughton, C. 1998, \apjs, 119, 121
\bibitem[Majewski et al.(2003)]{mswo03} Majewski, S. R., Skrutskie, M. F., Weinberg, M. D., and Ostheimer, J. C. 2003, \apj, 599, 1082
\bibitem[Martinez-Delgado et al.(2006)]{mpjai06} Martinez-Delgado, D., Penarrubia, J., Juric, M., Alfaro, E. J., Ivezic, Z. 2007, Ap. J. Suppl., 660, 1264
\bibitem[Martinez-Delgado et al.(2004)]{mgac04} Martinez-Delgado, D., Gomez-Flechoso, M. A., Aparicio, A., Carrera, R. 2004, \apj, 601,242
\bibitem[Moulataka, et al.(2004)]{mipc04} Moultaka, J., Ilovaisky, S. A., Prugniel, P., \& Soubiran, C. 2004, PASP, 116, 693
\bibitem[Monaco et al.(2006)]{metal06} Monaco, L., Bellazzini, M., Bonifacio, P., Buzzoni, A., Ferraro, F. R., Marconi, G., Sbordone, L., \& Zaggia, S. 2006, A\&A, 464, 201
\bibitem[Newberg, Yanny \& Willett(2009)]{cetuspolarstream} Newberg, H.J., Yanny, B. \& Willett, B.A. 2009, \apjl, submitted.
\bibitem[Newberg et al.(2007)]{netal07} Newberg, H., Yanny, B., Cole, N., Beers, T. , Re Fiorentin, P., Schneider, D., and Wilhelm, R. 2007, \apj, 668, 221.
\bibitem[Newberg \& Yanny(2006)]{ny06} Newberg, H., \& Yanny, B. 2006, in JPC Conf. Ser.: Physics at the end of the Galactic Cosmic Ray Spectrum, ed. G. Thomson \& P. Sokolsky, astro-ph/0507671
\bibitem[Newberg \& Yanny(2005)]{ny05} Newberg, H., \& Yanny, B. 2005, in ASP Conf. Ser. 338: Astrometry in the Age of the Next Generation of Large Telescopes, ed. P. K. Seidelmann \& A. K. B. Monet, 210, astro-ph/0502386
\bibitem[Newberg et al.(2003)]{netal03} Newberg, H., Yanny, B., et al. 2003, \apjl, 596, L191
\bibitem[Newberg et al.(2002)]{netal02} Newberg, H., Yanny, B., et al. 2002, \apj, 569, 245 
\bibitem[Odenkirchen et al.(2003)]{oetal03} Odenkirchen, M. et al. 2006, \aj, 126, 2385 
\bibitem[Parker, Humphreys, \& Larsen(2003)]{phl03} Parker, J. E., Humphreys, R. M., \& Larsen, J. A. 2003, \aj, 126, 1346
\bibitem[Pier et al.(2002)]{pmhhkli03} Pier, J. R., Munn, J. A., Hindsley, R. B., Hennessy, G. S., Kent, S. M., Lupton, R. H., and Ivezi\'{c}, Z. 2003, \aj, 125, 1559
\bibitem[Savage et al.(2006)]{snfg06} Savage, C., Newberg, H. J., Freese, K., \& Gondolo, P. 2006, Journal of Cosmology and Astroparticle Physics, 7, 3
\bibitem[Schlegel, Finkbeiner, \& Davis(1998)]{sfd98} Schlegel, D.J., Finkbeiner, D.P., \& Davis, M. 1998, ApJ, 500, 525
\bibitem[Searle \& Zinn(1978)]{1978ApJ...225..357S} Searle, L., \& Zinn, R.\ 1978, \apj, 225, 357 
\bibitem[Skrutskie et al.(2006)]{skrutskie06} Skrutskie, M.~F., et al.\ 2006, \aj, 131, 1163
\bibitem[Smith et al.(2002)]{setal02} Smith, J. A. et al. 2002, \aj, 123, 2121
\bibitem[Stoughton et al.(2001)]{setal01} Stoughton, C., et al. 2001, \aj, 123, 485
\bibitem[Tucker et al.(2006)]{tetal06} Tucker, D., et al. 2006, Astronomische Nachrichten, 325, 583
\bibitem[Vivas et al.(2001)]{vetal01} Vivas, A.~K.~et al.\ 2001, \apjl, 554, L33
\bibitem[Vivas, Zinn, \& Gallart(2005)]{vzg05} Vivas, A. K., Zinn, R. \& Gallart, C. 2005, AJ, 129, 189.
\bibitem[Wilhelm, Beers, \& Gray(1999)]{wbg99} Wilhelm, R., Beers, T. C., \& Gray, R. O. 1999, \aj, 117, 2308
\bibitem[Willett et al.(2009)]{2009arXiv0901.4046W} Willett, B.~A., 
Newberg, H.~J., Zhang, H., Yanny, B., 
\& Beers, T.~C.\ 2009, arXiv:0901.4046 
\bibitem[Xu, Deng \& Hu(2006)]{xdh06} Xu, Y., Deng, L. C. \& Hu, J. Y. 2006, \mnras, 368, 1811
\bibitem[Xu, Deng \& Hu(2007)]{xdh07} Xu, Y., Deng, L. C. \& Hu, J. Y. 2007, \mnras, 379, 1373.
\bibitem[Yanny et al.(2009)]{yetal09} Yanny, B., et al.\ 2009, \aj, 137, 4377.
\bibitem[Yanny et al.(2000)]{ynetal00} Yanny, B., Newberg, H. J., et al. 2000, \apj, 540, 825
\bibitem[Yanny et al.(2003)]{yetal03} Yanny, B., Newberg, H. J., et al. 2003, \apj, 588, 841  
\bibitem[York et al.(2000)]{yetal00} York, D.G.  et al. 2000, \aj, 120, 1579
\bibitem[Zinn et al.(2004)]{zvgw04} Zinn, R., Vivas, A. K., Gallart, C. \& Winnick, R. 2004, ASP Conf. Series, 327, 92
\end{thebibliography}
\end{document}